\documentclass{natureprintstyle}
\usepackage{graphicx}
\usepackage{longtable}
\usepackage{supertabular}

\pdfoutput=1

\def\aap{Astron. Astrophys.}                
\def\aas{Astron. Astrophys. Supp.}  
\def\apj{Astrophys. J.}                 
\def\aj{Astron. J.}                   
\def\apjl{Astrophys. J.}                
\def\araa{Ann. Rev. Astron. Astrophys.}             

\def\apjs{Astrophys. J. Supp.}               
\def\mnras{Mon. Not. R. Astron. Soc.}             
\def\nat{Nature}
\def\pasp{Pub. Astron. Soc. Pacific}  
\def\spie{Proc. Soc. Photo-optical Instrumentation Engineers}  
\def\beq{\begin{equation}}
\def\eeq{\end{equation}}
\newcommand{\kms}{\,{\rm km\,s^{-1}}}
\newcommand{\msun}{\,M_{\odot}}
\newcommand{\lsun}{L_{\odot}}
\newcommand{\mbh}{M_{\rm BH}}
\newcommand{\mlstars}{M_\star/L}

\title{Two ten-billion-solar-mass black holes at the centres of giant elliptical galaxies}

\author{Nicholas J. McConnell$^1$,  Chung-Pei Ma$^1$,  Karl Gebhardt$^2$,  Shelley A. Wright$^1$, Jeremy D. Murphy$^2$,  Tod R. Lauer$^3$,  James R. Graham$^{1,4}$ and Douglas O. Richstone$^5$}

\begin{document}

\maketitle

\begin{affiliations}
\item Department of Astronomy, University of California, Berkeley, California 94720, USA.
\item Department of Astronomy, University of Texas, Austin, Texas 78712, USA.
\item National Optical Astronomy Observatory, Tucson, Arizona 85726, USA.
\item Dunlap Institute for Astronomy and Astrophysics, University of Toronto, Ontario, Canada.
\item Department of Astronomy, University of Michigan, Ann Arbor, Michigan 48109, USA.
\end{affiliations}

\begin{abstract}

  Observational work conducted over the last few decades 
  indicates that all massive galaxies
  have supermassive black holes at their centres.
  Although the luminosities and brightness fluctuations of quasars in the
  early Universe suggest that some are powered by black holes with masses greater than 10 billion solar masses \cite{netzer03,vestergaard08}, the remnants of these objects have not
  been found in the nearby Universe.  The giant
  elliptical galaxy Messier 87 hosts the hitherto most massive known black
  hole, which has a mass of 6.3 billion solar masses \cite{sargent78,gebhardt11}.  Here we report that
 NGC 3842, the brightest galaxy in a cluster at a distance from Earth of 98 megaparsecs, has a central black hole with a mass of 9.7 billion solar masses, and that a 
black hole of comparable or greater mass is present in NGC 4889, the
  brightest galaxy in the Coma cluster (at a distance of 103 megaparsecs).  
  These two black holes are significantly more massive
  than predicted by linearly extrapolating the widely-used correlations
  between black hole mass and the stellar velocity dispersion or bulge
  luminosity of the host galaxy \cite{dressler89,kormendy95,fm00,gebhardt00,gultekin09}.  
Although these correlations remain useful
  for predicting black hole masses in less massive elliptical galaxies, our measurements suggest that 
  different evolutionary processes influence the growth of the largest
  galaxies and their black holes.

\end{abstract}

Empirical scaling relations between black hole mass ($\mbh$), galaxy bulge velocity
dispersion ($\sigma$), and luminosity ($L$) are commonly used to estimate
black hole masses, because for most galaxies we are unable to make a direct measurement.
Estimates of the number density of black holes in a given mass range thus
depend upon the empirically determined $\mbh -\sigma$ and
$\mbh - L$ relations over an appropriate range of galaxy masses.  Directly
measuring $\mbh$ from the kinematics of stars or gas in the vicinity of
the black hole is particularly difficult at the highest galaxy masses, because massive galaxies are rare, their typical distances from Earth are large, and their central stellar densities are relatively low.  The most massive galaxies are typically brightest cluster galaxies (BCGs), that is,
giant ellipticals that reside near the centres of galaxy clusters.

We have obtained high-resolution, two-dimensional data of the line-of-sight
stellar velocities in the central regions of NGC 3842 and NGC 4889 using
integral field spectrographs at the Gemini North and Keck 2 telescopes, in Hawaii.
The stellar luminosity distribution of each galaxy is provided by surface
photometry from NASA's Hubble Space Telescope and ground-based telescopes
\cite{laine03,PL95}.  NGC 3842 is the BCG of Abell 1367, a moderately rich
galaxy cluster.  NGC 4889 is the BCG of the Coma cluster (Abell 1656), one
of the richest nearby galaxy clusters.  We targeted these two galaxies
because they have relatively high central surface brightnesses and lie at an
accessible distance for direct measurements of $\mbh$.

\begin{figure}[b!]
\begin{center}
\includegraphics[width=3.8in]{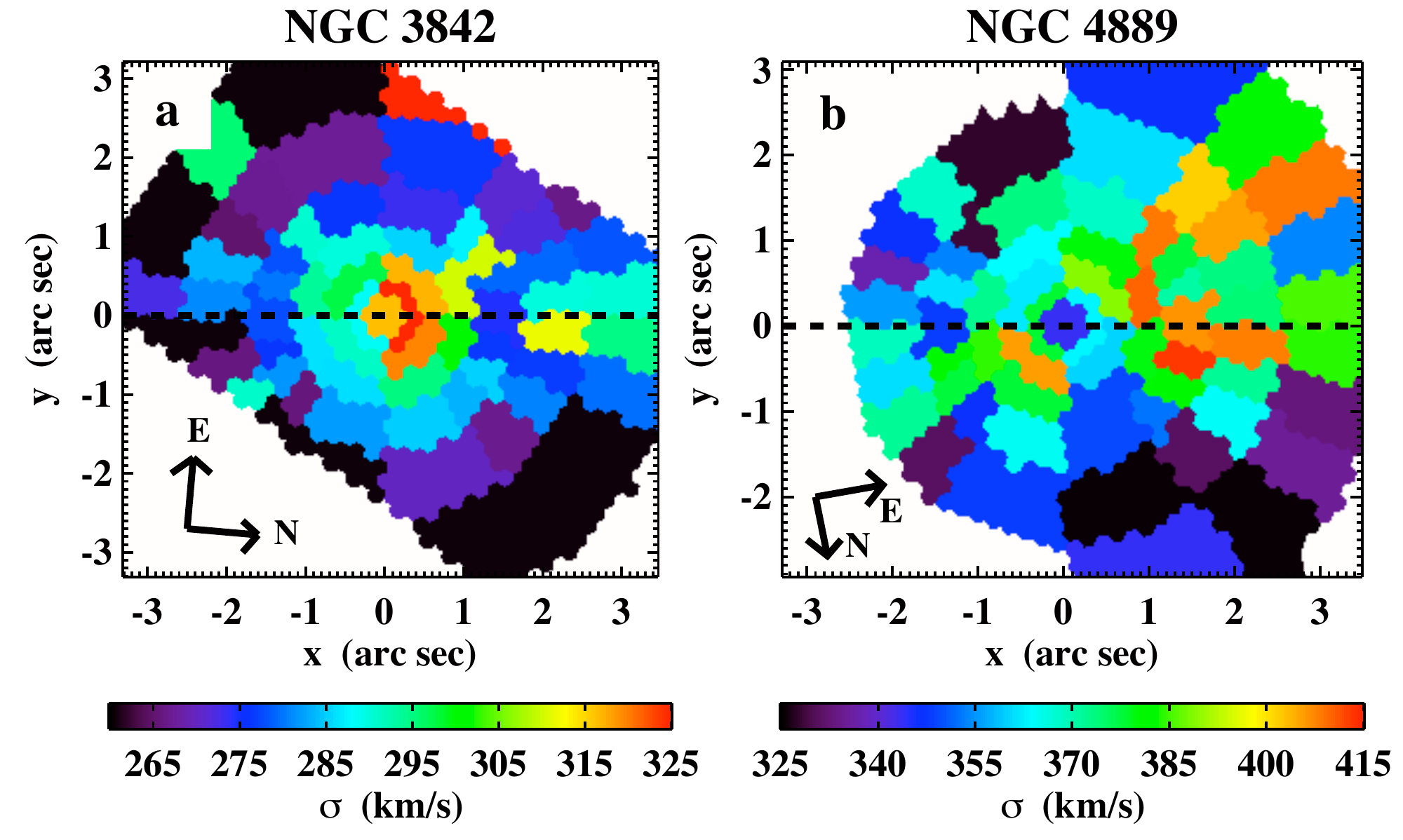}
\end{center}
\vspace{-0.7cm}
\caption{{\bf Two-dimensional maps of the line-of-sight stellar velocity
  dispersions.}  The maps show the central regions of NGC 3842 (\textbf{a})
and NGC 4889 (\textbf{b})
observed using the GMOS spectrograph \cite{allington} on the 8-m Gemini North telescope.
  Additional kinematics at large radii were measured using the VIRUS-P spectrograph \cite{hill} at the 2.7-m Harlan J. Smith
telescope, and additional high-resolution data were acquired with the OSIRIS spectrograph \cite{larkin} at the 10-m Keck 2 telescope.  GMOS, OSIRIS, and VIRUS-P are all integral field spectrographs, which record spectra at multiple positions in a two-dimensional spatial array. 
The horizontal dashed line in each panel traces the major axis of the
galaxy.  The median 1 s.d. errors in velocity dispersion are $12 \kms$ and $20 \kms$ for NGC
3842 and NGC 4889, respectively. 
In NGC 4889 the highest velocity dispersions, near $410 \kms$, are located
on the east side of the galaxy, at least 1.1 arcsec from the centre.  }
\label{fig:2d}
\end{figure}
\begin{figure}[b!]
\vspace{3.5in}
\begin{center}
\includegraphics[width=3.1in]{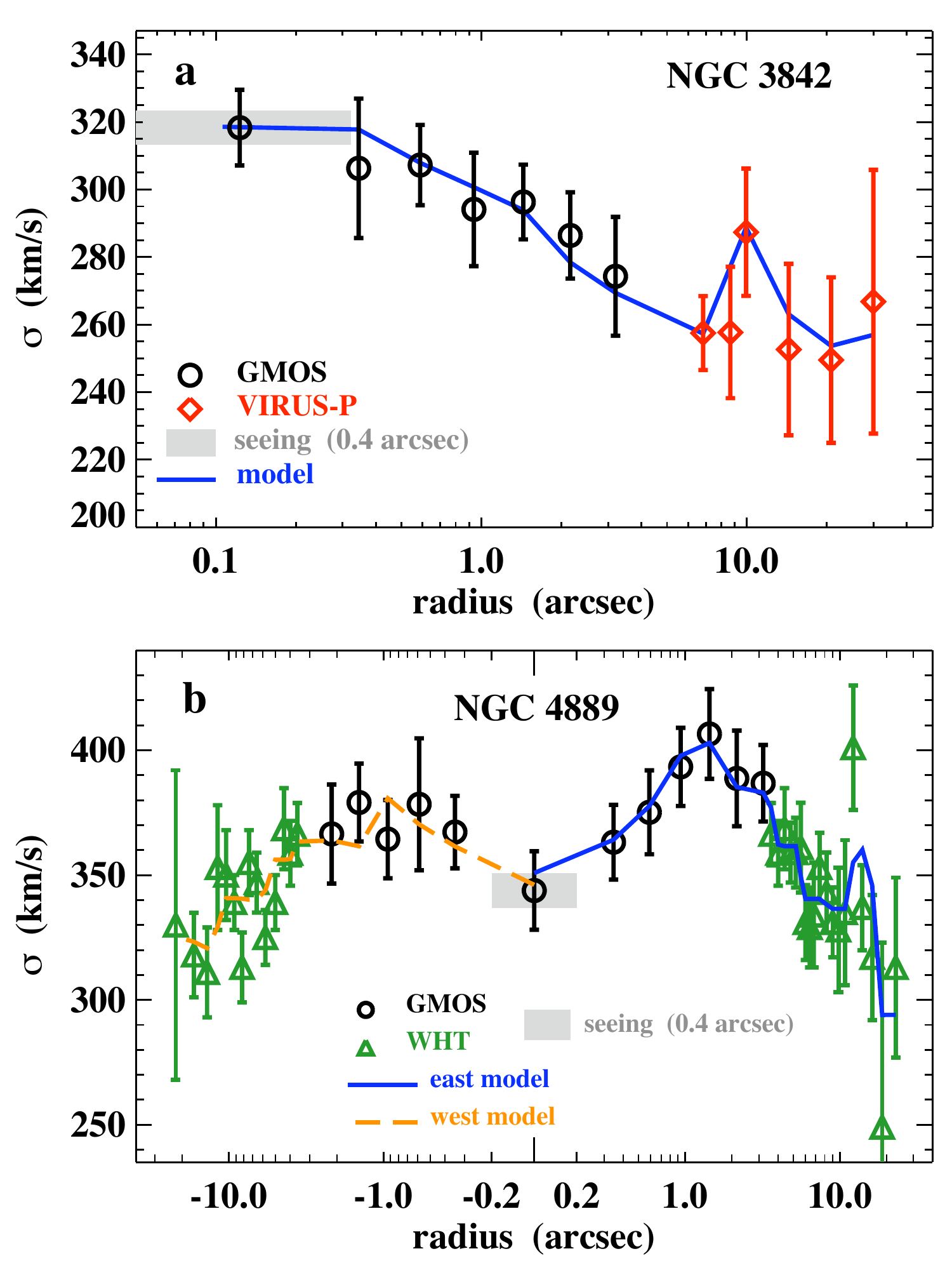}
\end{center}
\vspace{-0.2in}
\caption{{\bf One-dimensional profiles of the line-of-sight velocity dispersions.}  \textbf{(a)} Dispersion versus radius in NGC 3842, after averaging data at a given radius, based on measurements with GMOS (black circles) and VIRUS-P (red diamonds).  The solid blue line is the projected line-of-sight dispersion from our best-fitting stellar orbit model of NGC 3842.  
\textbf{(b)}  Dispersion versus radius along the major axis of NGC 4889, measured from GMOS (black circles) and the William Herschel Telescope \cite{loubser08} (green diamonds).  The maximum velocity dispersion occurs at $r = 1.4$ arcsec.  The solid blue line is the projected line-of-sight dispersion from our best-fitting orbit model using data from the east side of NGC 4889 ($r > 0$).  The dashed orange line is from our best-fitting orbit model using data from the west side of NGC 4889 ($r < 0$).  Error bars, 1 s.d.}
\label{fig:1d}
\end{figure}

We measured the distribution of stellar velocities at 82 different
locations in NGC 3842.  The line-of-sight velocity dispersion in NGC 3842 is
between 270 and $300 \kms$ at large galactocentric radii ($r$) and rises in the central 0.7
arcsec ($r < 330$ pc), peaking at $326 \kms$ (Figs
~\ref{fig:2d}-\ref{fig:1d}).  
We determined the mass of the central black hole by constructing
a series of orbit superposition models \cite{schwarz}.  Each model assumes
a black hole mass, stellar mass-to-light ratio ($\mlstars$) and dark matter profile,
and generates a library of time-averaged stellar orbits in the resulting
gravitational potential.  The model then fits a weighted combination of
orbital line-of-sight velocities to the set of measured stellar velocity
distributions.  The goodness-of-fit statistic $\chi^2$ is computed as a
function of the assumed values of $\mbh$ and the stellar mass-to-light ratio.
Using our best-fitting model dark matter halo,
we measure a black hole mass of $9.7 \times 10^9$ solar masses ($\msun$), with a 68\%
confidence interval of 7.2 - $12.7\times 10^9 \msun$.  Models with no black
hole are ruled out at the 99.996\% confidence level ($\Delta \chi^2 =
17.1$).  We find the stellar mass-to-light ratio to equal $5.1
\msun/\lsun$ in $R$ band ($\lsun$, solar luminosity), with a 68\% confidence interval of $4.4 \msun/\lsun - 5.8 \msun/\lsun$).

We measured stellar velocity distributions at 63 locations in NGC 4889 and combined our measurements
with published long-slit kinematics at larger radii \cite{loubser08}.  The
largest velocity dispersions in NGC 4889 are located across an extended
region on the east side of the galaxy.  The stellar orbits in our models
are defined to be symmetric about the galaxy centre, so we constrain $\mbh$
by running separate trials with velocity profiles from four quadrants of
the galaxy.  
The best-fitting black hole masses from the four quadrants range from $9.8 \times 10^9 \msun$ to $2.7 \times 10^{10} \msun$.    
All quadrants favor tangential orbits near the galaxy centre, which cause the line-of-sight velocity dispersion to decrease even as the internal three-dimensional velocity dispersion increases toward the black hole.
Although no single model is consistent with all of the observed kinematic
features in NGC 4889, we can define a confidence interval for $\mbh$ by
considering the most extreme confidence limits from the cumulative set of
models.  The corresponding 68\% confidence interval is $(0.6 - 3.7) \times
10^{10} \msun$.  We adopt a black hole mass of $2.1 \times 10^{10} \msun$,
corresponding to the midpoint of this interval.

\begin{figure*}[h!]
\includegraphics[width=7.6in]{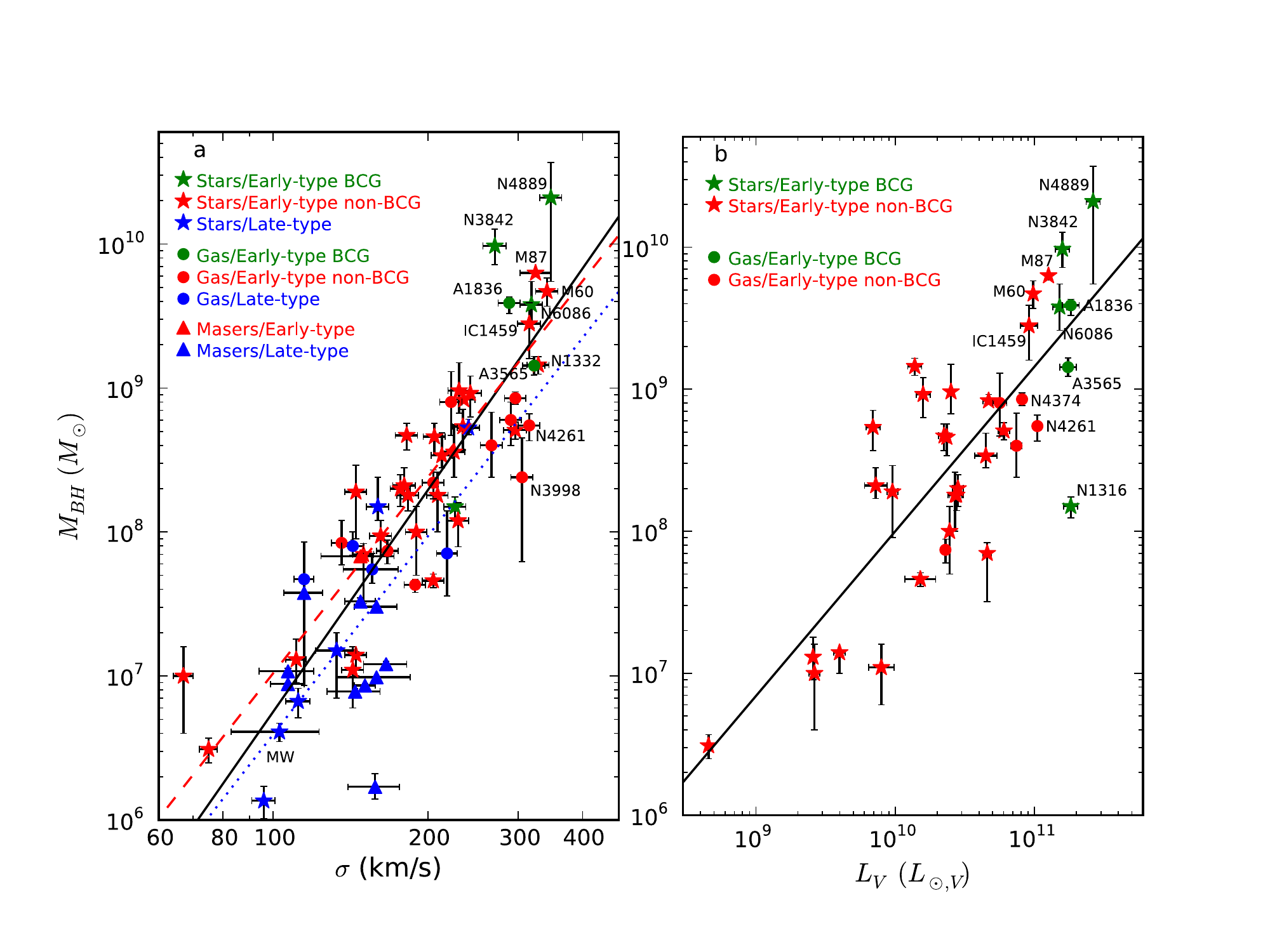}
\caption{{\bf Correlations of dynamically measured black hole masses and
    bulk properties of host galaxies.}  \textbf{(a)} Black hole mass,
  $\mbh$, versus stellar velocity dispersion, $\sigma$, for 65 galaxies
  with direct dynamical measurements of $\mbh$.  For galaxies with spatially resolved stellar kinematics, $\sigma$ is the luminosity-weighted average within one
  effective radius (Supplementary Information).  \textbf{(b)} Black hole mass versus $V$-band bulge
  luminosity, $L_V$ ($L_{\odot, V}$, solar value), for 36 early-type galaxies with direct dynamical
  measurements of $\mbh$.  Our sample of 65 galaxies consists of 32
  measurements from a 2009 compilation \cite{gultekin09}, 16 galaxies with
  masses updated since 2009, 15 new galaxies with $\mbh$ measurements, and
  the two galaxies reported here.  A complete list of the galaxies is given
  in Supplementary Table 4.  BCGs (defined here as the most luminous galaxy in a cluster) are plotted
  in green, other elliptical and S0 galaxies are plotted in red, and
  late-type spiral galaxies are plotted in blue.  
     The black hole masses are measured using dynamics of masers (triangles), stars (stars), or gas (circles).
   Error bars, 68\% confidence intervals.  For most
  of the maser galaxies, the error bars in $\mbh$ are smaller than the
  plotted symbol.  The solid black line in \textbf{(a)}
  shows the best-fitting power law for the entire sample:
  $\log_{10}(\mbh/\msun) = 8.29 + 5.12 \log_{10}[\sigma / (200 \kms)]$.
 When early-type and late-type galaxies are fit separately,
  the resulting power laws are $\log_{10}(\mbh/\msun) = 8.38 + 4.53
  \log_{10}[\sigma /(200 \kms)]$ for elliptical and S0 galaxies (dashed red
  line), and $\log_{10}(\mbh/\msun) = 7.97 + 4.58 \log_{10}[\sigma / (200
  \kms)]$ for spiral galaxies (dotted blue line).  
  The solid black line in \textbf{(b)} shows the best-fitting power law:
  $\log_{10}(\mbh/\msun) = 9.16 + 1.16 \log_{10}(L_V / 10^{11} \lsun)$. 
  We do not label Messier 87 as a BCG, as is commonly done, as NGC 4472 in the Virgo cluster is 0.2 mag brighter.  
}
\label{fig:bhfits}
\end{figure*}

Figure~\ref{fig:bhfits} shows the $\mbh-\sigma$ and $\mbh-L$ relations,
using data compiled from studies published before the end of August 2011, plus
our measurements of NGC 3842 and NGC 4889.  Tabulated data with references
are provided as Supplementary
Information.  The most widely used form for both relations is a power law
with a constant exponent.  Straight lines in Fig.~\ref{fig:bhfits} show our fits to $\mbh(\sigma)$ and $\mbh(L)$.  The relationship between $\sigma$ and $L$, however,
flattens at high galaxy masses, and constant-exponent power laws for
the $\mbh-\sigma$ and $\mbh-L$ relations produce contradictory predictions for $\mbh$ in
this mass range \cite{lauer07}.  Direct measurements of $\mbh$ in higher
mass galaxies will compel the revision of one or both of the $\mbh-\sigma$ and
$\mbh-L$ relations.

The average velocity dispersion in NGC 3842 is $270 \kms$, measured outside the black hole radius of influence (1.2 arcsec or 570 pc) and inside the two-dimensional half-light radius (38 arcsec or 18 kpc).  Although NGC 3842 hosts a black hole more massive than any previously detected, its average dispersion
ranks only fourteenth among 65 galaxies with direct measurements of $\mbh$.
Its luminosity ranks fifth in this sample of galaxies and is exceeded only by other
BCGs.  On the basis of $\sigma$ and $L$ for NGC 3842, our revised $\mbh-\sigma$ and $\mbh -
L$ relations predict $\mbh = 9.1 \times 10^8 \msun$ and $2.5 \times
10^9 \msun$, respectively.  Similarly, for
NGC 4889 the respective predictions are $3.3 \times 10^9 \msun$ and $4.5 \times 10^9 \msun$.  These predictions are smaller than our direct
measurements of $\mbh$, by 1.6-4.6 times the 1-s.d. scatter in
the $\mbh-\sigma$ and $\mbh-L$ relations \cite{gultekin09}.
Four measurements of $\mbh$ in BCGs existed before this work.
Two measurements based on gas dynamics and one based on stellar dynamics
all lie within 1.2 s.d. of our revised fits to the $\mbh-\sigma$ and
$\mbh-L$ relations \cite{dallabonta09,mcconnell11}.  Yet the measurement of $\mbh$ in NGC 1316, the
BCG of the Fornax cluster, is 3.4 s.d. less than that predicted by our $\mbh-L$ relation \cite{nowak08}.
The high scatter indicated by this collection of measurements reveals large
uncertainties in the standard practice of using galacitc $\sigma$ or $L$ as a proxy for the central black hole mass in giant
elliptical galaxies and their predecessors.

Several BCGs within 200 Mpc of Earth are at least twice as luminous as NGC 3842, and
three times as luminous as M87, which hosted the most massive black hole known before this work.  
In spite of their extreme luminosities, BCGs have velocity dispersions similar to those of the most massive field elliptical galaxies.  Yet the most massive black holes are found predominantly in BCGs (Fig.~\ref{fig:bhfits}).
How galaxies are assembled and the role of gas dissipation 
affect the correlations (or lack
thereof) among $\mbh$, $\sigma$, and $L$.  
Simulations of mergers of gas-rich disk galaxies are able to produce
remnant galaxies that follow the observed $\mbh-\sigma$ correlation 
in Fig. 3a, over the intermediate mass range $\mbh \approx 10^7 - 10^9 \msun$
(refs 18, 19). 
By contrast, simulated mergers of elliptical galaxies
with low-angular momentum progenitor orbits increase $\mbh$
and $L$ by similar numerical factors, without increasing the velocity
dispersion \cite{boylan06}.  Because these mergers are a likely path to forming the most massive galaxies, the $\mbh-\sigma$ correlation may steepen or disappear altogether at the highest galaxy masses.  Massive
elliptical galaxies retain residual quantities of gas even after the
decline of star formation.  Accretion of this gas onto the galaxies'
central black holes could help increase $\mbh$ and further steepen the
$\mbh-\sigma$ and $\mbh-L$ relations.  

Black holes in excess of $10^{10} \msun$ are observed as quasars in the
early Universe, from $1.4 \times 10^9$ to $3.3 \times 10^9$ yr after the Big Bang \cite{vestergaard08} (redshift, $z = 2-4.5$).
Throughout the last $1.0 \times 10^{10}$ yr, however, these extremely massive
black holes have not been accreting appreciably, and the average mass of the black holes
powering quasars has decreased steadily.  Quasar
activity and elliptical galaxy formation are predicted to arise from
similar merger-triggered processes, and there is growing evidence that
present-day massive elliptical galaxies once hosted the most luminous high-redshift
quasars \cite{hopkins07}.  Yet definitive classification of these
quasars' host galaxies has remained elusive.

Our measurements of black holes with masses of around $10^{10} \msun$ in NGC 3842 and
NGC 4889 provide circumstantial evidence that BCGs host the remnants of
extremely luminous quasars.   
The number density of nearby BCGs ($\sim 5 \times 10^{-6}$ Mpc$^{-3}$) 
is consistent with 
 the number density of black holes ($\sim 3\times 10^{-7}$
to $10^{-5}$ Mpc$^{-3}$) with masses between $10^9 \msun$ and $10^{10} \msun$
predicted from the $\mbh-L$ relation and the luminosity function of nearby galaxies.  Furthermore, both quantities agree with predictions based on the black hole masses and duty cycles of quasars.  
The black hole number density predicted from the
$\mbh-\sigma$ relation, however, is an order of magnitude less than the inferred
quasar population \cite{lauer07, lauer07b}.  These two predictions can be reconciled if
the $\mbh-\sigma$ relation has upward curvature or a large degree of
intrinsic scatter in $\mbh$ at the high-mass end, as suggested by our new measurements.  
With improvements in adaptive optics instrumentation on
large optical telescopes and very-long baseline interferometry at radio
wavelengths, black holes are being sought and detected in increasingly
exotic host galaxies.  Along with our measurements of the black hole masses
in NGC 3842 and NGC 4889, future measurements in other 
massive galaxies will quantify
the cumulative growth of supermassive black holes in the Universe's densest environments.

\begin{addendum}
\item[Supplementary Information] is linked to the online version of the paper at
http://www.nature.com/nature.

 \item[Acknowledgments] 
   N.J.M., C.-P.M, K.G. and J.R.G. are supported by the National Science Foundation.   C.-P.M. is supported by NASA
   and by the Miller Institute for Basic Research in Science, University of
   California, Berkeley.  S.A.W. is
   supported by NASA through the Hubble Fellowship.
   Data presented here were obtained using Gemini Observatory, W.M. Keck
   Observatory, and McDonald Observatory.  The Gemini Observatory is operated by the
   Association of Universities for Research in Astronomy, Inc., under a
   cooperative agreement with the National Science Foundation on behalf of the Gemini partnership.
   The W. M. Keck Observatory is operated as a scientific partnership among
   the California Institute of Technology, the University of California,
   and NASA.  The McDonald
   Observatory is operated by the University of Texas at Austin.  The instrument VIRUS-P was funded by G. and C. Mitchell.  Stellar
   orbit models were run using the facilities at the Texas Advanced
   Computing Center at The University of Texas at Austin.
   

\item[Author Contributions] 

  N.J.M. carried out the data analysis and modeling.
 N.J.M, C.-P.M. and S.A.W. wrote the manuscript.
C.-P.M. compiled the data for Figure 3 and
  oversaw communication among co-authors.  S.A.W. analyzed OSIRIS
  data of NGC 3842.  K.G. provided GMOS data of NGC 3842 and NGC 4889.
  K.G. and D.O.R. developed the stellar orbit modeling code.
  J.D.M. provided VIRUS-P data of NGC 3842.  T.R.L. provided photometric
  data and image analysis of NGC 3842 and NGC 4889.  J.R.G. led the OSIRIS
  observing campaign for NGC 3842.  All authors contributed to the
  interpretive analysis of the observations and the writing of the paper.

\item[Competing Interests] The authors declare that they have no
competing financial interests.

 \item[Correspondence] Correspondence and requests for materials
should be addressed to N. McConnell (email: nmcc@berkeley.edu) or C.-P. Ma (cpma@berkeley.edu).

\end{addendum}

\clearpage

\begin{center}
\title  {\huge \bf{Supplementary Information}}
\end{center}
\author{}
\date{}

In the first section, we describe spectroscopic data of NGC 3842 and NGC 4889 and our procedures for measuring stellar kinematics.  In the second section, we describe photometric data.  In the third section we compare overlapping kinematic measurements from GMOS, OSIRIS, VIRUS-P, and long-slit spectroscopy.  The fourth section contains a summary of our stellar orbit models.  In the fifth section we describe the unusual kinematics and stellar orbital structures in NGC 4889, and justify our measurement of $\mbh$.  In the final section we describe our updated sample and fits to the $\mbh - \sigma$ and $\mbh - L$ relations.

We tabulate our spectroscopic observations, our results from modeling each quadrant of NGC 4889, our fits to the $\mbh - \sigma$ and $\mbh - L$ relations, and the galaxy sample used to fit the relations.  Our measured line-of-sight velocity distributions (LOSVDs) for NGC 3842 and NGC 4889 are available as online data.

\section*{Spectroscopic Data and LOSVD Extraction\\}

We map stellar orbital motions in NGC 3842 and NGC 4889 by measuring the
LOSVD for different regions in each
galaxy. Each LOSVD is determined by fitting a composite template stellar
spectrum to a fully reduced spectrum of the galaxy.  The LOSVDs are
non-parametric probability distributions, defined at each of 15 velocity
bins. We use a Maximum Penalized Likelihood technique to optimize the
LOSVD value in each velocity bin while simultaneously optimizing the
weights of individual template stars \cite{geb00aj, pinkney03, nowak08}.  

\begin{figure}[b!]
\begin{center}
\includegraphics[width=3.3in]{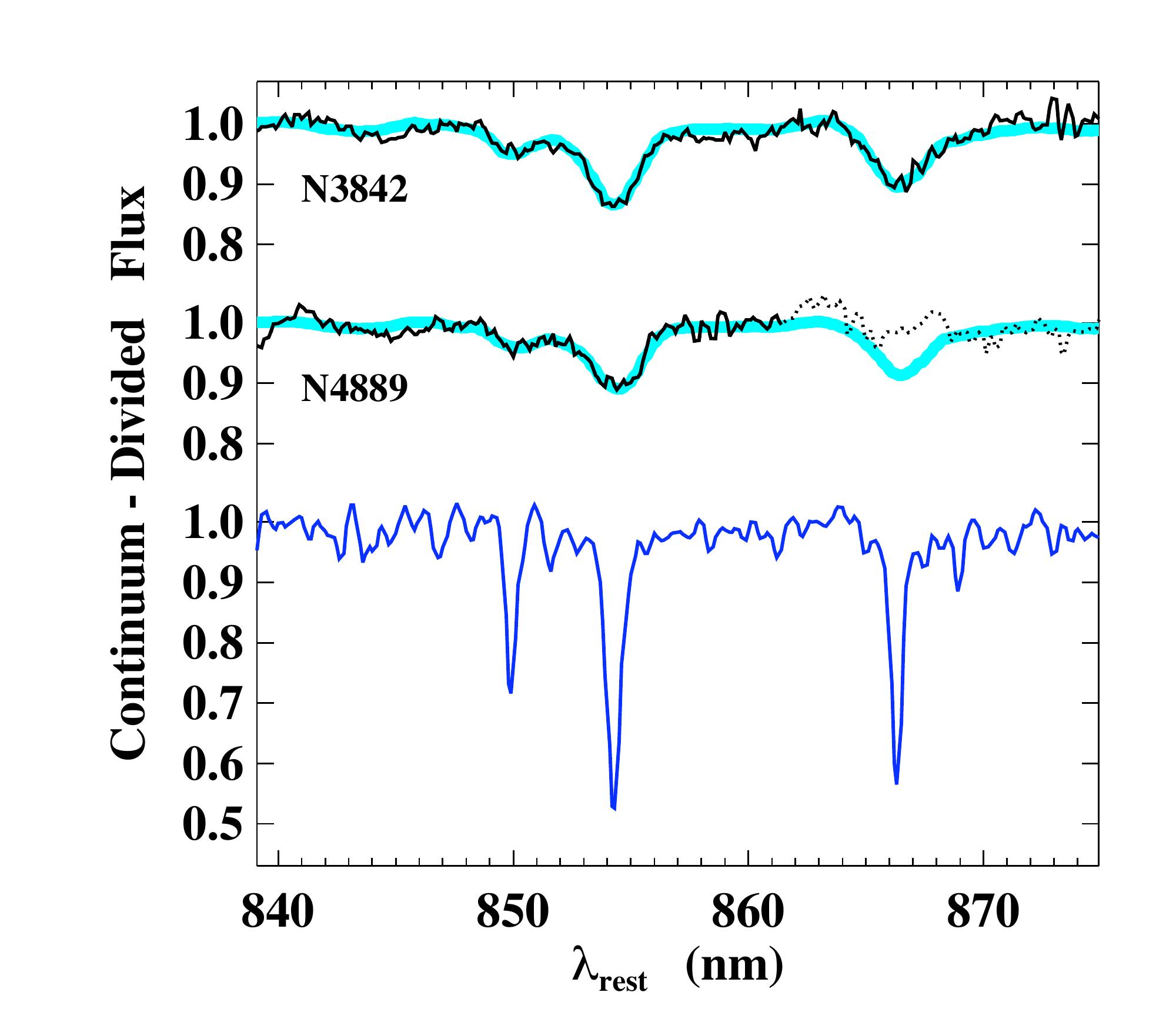}
\end{center}
\vspace{-0.7cm}
\caption{GMOS spectra of NGC 3842 and NGC 4889.  Each spectrum corresponds to the center of the galaxy ($r < 0.25$ arcseconds).  The upper spectrum is NGC 3842 (black), overlaid with the best-fitting, LOSVD-convolved template spectrum (thick, light blue).  The middle spectrum is NGC 4889, overlaid with the best-fitting template spectrum.  The dotted portion of the spectrum was excluded from the LOSVD fitting.  The lower spectrum is template star HD 73710 (G9III), before convolution with the LOSVD.}
\label{fig:gmos}
\end{figure}

\begin{figure}[b!]
\begin{center}
\includegraphics[width=3.8in]{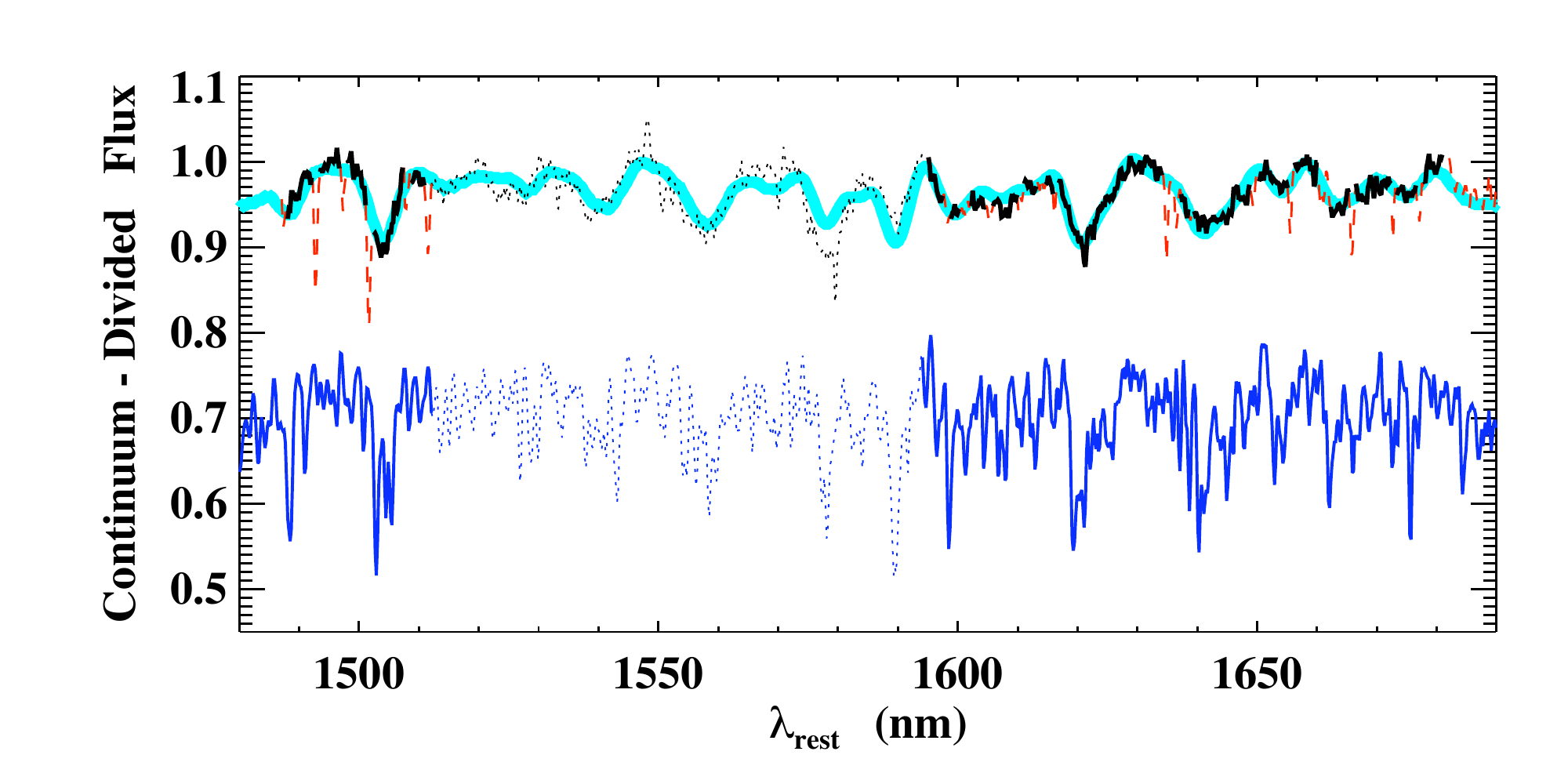}
\end{center}
\vspace{-0.7cm}
\caption{OSIRIS spectrum of the central region of NGC 3842 ($r < 0.25$ arcseconds, or 110 pc).  The upper spectrum is NGC 3842 (black), overlaid with the best-fitting, LOSVD-convolved template spectrum (thick, light blue).  The red dashed lines in the galaxy spectrum are residuals from imperfectly subtracted sky emission.  The lower spectrum is the best-fit composite template before convolution with the LOSVD.  Our observed template stars fit spectra of NGC 3842 poorly across the dotted region from 1510 to 1590 nm, and therefore this region is excluded from kinematic fitting.}
\label{fig:osiris}
\end{figure}

\begin{figure}[b!]
\begin{center}
\includegraphics[width=3.8in]{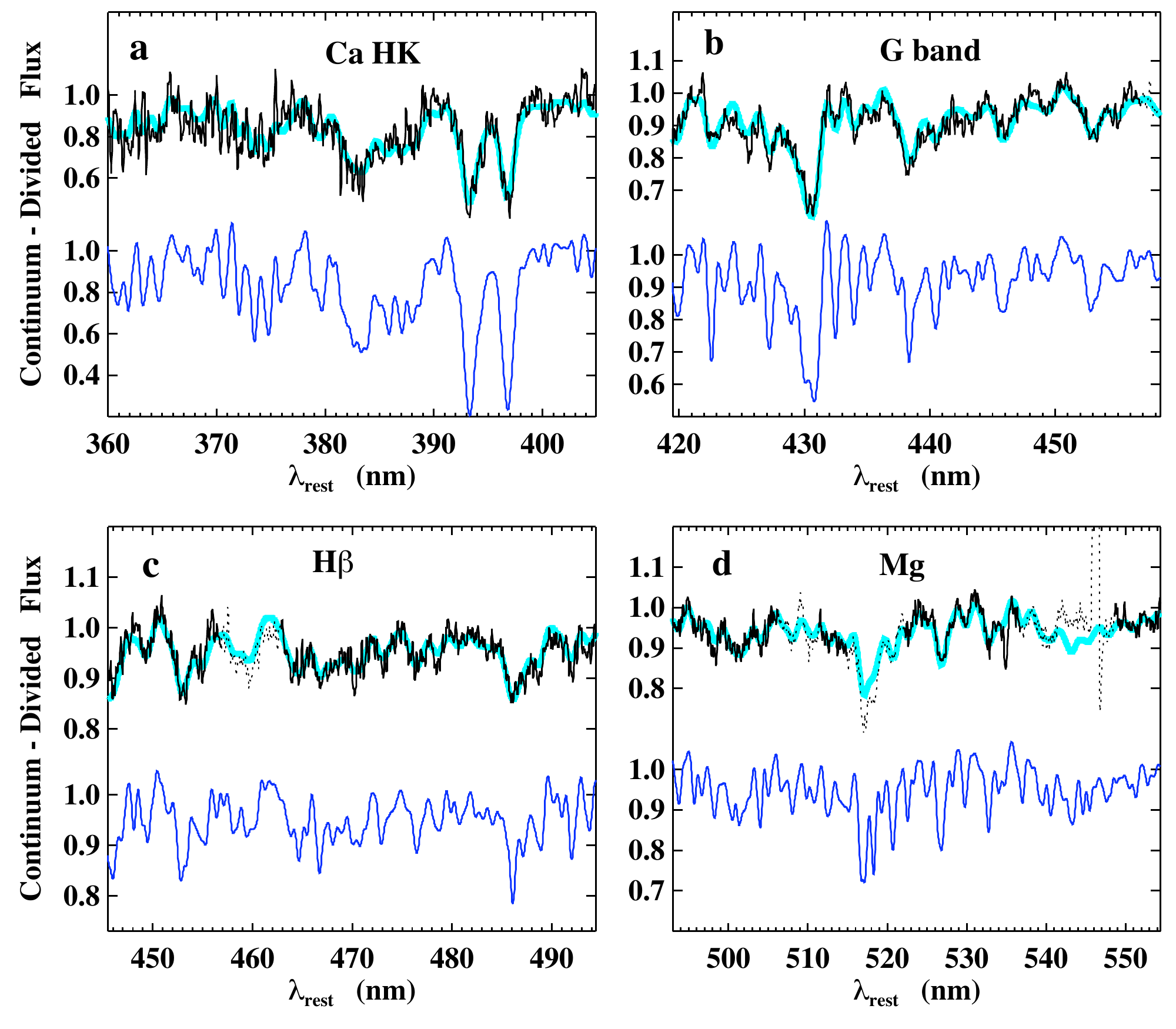}
\end{center}
\vspace{-0.7cm}
\caption{VIRUS-P spectrum of NGC 3842, corresponding to a semi-annulus with an inner radius of 17.0 arcseconds (7.7 kpc) and an outer radius of 24.5 arcseconds (11.0 kpc).  The four panels contain different sub-regions of the galaxy spectrum.  Each sub-region is evaluated independently for a best-fit LOSVD and best-fit composite template spectrum. The upper spectrum in each panel is NGC 3842 (black), overlaid with the best-fitting, LOSVD-convolved template spectrum (thick, light blue).  Dotted portions of the galaxy spectrum have been masked from the fit.  The lower spectrum in each panel is the best-fit composite template before convolution with the LOSVD.}
\label{fig:virus}
\end{figure}

In Table~\ref{tab:spec}, we summarize our observations with the integral field spectrographs GMOS, OSIRIS, and VIRUS-P.  Figures~\ref{fig:gmos}-\ref{fig:virus} illustrate a sample galaxy and template spectrum from each instrument.  Our data from GMOS only cover radii within 3.8 arcseconds (1.8 kpc) of the center of NGC 3842 and NGC 4889, and by themselves cannot fully remove the degeneracies between $\mbh$ and $\mlstars$.  For NGC 3842, our VIRUS-P measurements cover radii out to 35.3 arcseconds (16.8 kpc) and can distinguish the enclosed stellar mass profile from the galaxy's dark matter halo.  This allows for an accurate determination of $\mlstars$, such that the GMOS data can accurately constrain $\mbh$.  
At radii from 3.6 to 23.0 arcseconds ($1.8-11.5$ kpc) along the major axis of NGC 4889, we use Gaussian velocity profiles from Loubser et al. (2008) \cite{loubser08}.

Our GMOS spectra for NGC 3842 and NGC 4889 are centered on the calcium triplet
absorption lines near 860 nm.  A sample GMOS spectrum for each galaxy is shown in
Figure~\ref{fig:gmos}, demonstrating the clean line profiles that are
typical for this spectral region. Another advantage to using the calcium
triplet is that kinematic measurements are not highly sensitive to the
stellar template used \cite{barth02}.  

OSIRIS data of NGC 3842 were acquired with the 0.05 arcsecond spatial scale and the
broad $H$-band filter, which spans a large number of atomic and molecular
absorption features. To measure kinematics, we fit carbon monoxide band
heads at 1598, 1619, 1640, and 1661 nm, and a deep magnesium feature near
1500 nm. 
The most severe source of noise in our OSIRIS spectra is residual
narrow-line emission from the night sky. This background emission varies
rapidly and is only partially corrected by recording 15-minute sky frames
in between pairs of 15-minute science exposures.  Even when masked, the
contaminated channels represent a non-negligible loss of spectral
information. This loss must be countered by increasing the signal-to-noise ratio
in the usable parts of the spectrum. To achieve adequate signal-to-noise,
we bin the data to the same
spatial regions as the overlapping LOSVDs from GMOS. Each of the final bins
contains approximately 80 OSIRIS spatial pixels.

VIRUS-P data were acquired in low-resolution mode, which provides
broad wavelength coverage. The poorest instrumental resolution over our
field is 0.56 nm full width at half-maximum (FWHM), which is adequate
considering that stellar velocity dispersions in NGC 3842 are typically
near $250 \kms$, corresponding to a resolution of approximately 0.7 nm FWHM $\times (\lambda$ / 360 nm). 
We fit VIRUS-P spectra with 16 template stars from the Indo-US library \cite{valdes04}, spanning spectral types from B9 to M3 and including stars
with sub-solar and super-solar metallicities.  
However, some spectral regions require additional adjustments to account for
metallicities and elemental abundance ratios outside the range of our
template library. We do not attempt to fit the entire spectral range
simultaneously, but instead follow the same fitting procedure used by Murphy et al. (2011) \cite{murphy11} for VIRUS-P data of M87. Each spectrum is divided
into four sub-regions, as illustrated in Figure~\ref{fig:virus}. We
independently determine the best-fit LOSVD for each sub-region and
discard any sub-regions that fail to produce a believable fit.  We then average the LOSVDs derived from individual spectral sub-regions.

Our stellar orbit models are axisymmetric, and so each of our final LOSVDs
must represent an average over four quadrants of the galaxy.  In order to preserve any rotational signal along the major axis, we invert the velocities from LOSVDs on the south side of NGC 3842 and the west side of NGC 4889.  However, neither galaxy shows strong rotation.  In NGC 3842, the resulting kinematics are sufficiently symmetric to average the LOSVDs from opposite sides of the galaxy.  
In NGC 4889, we have modeled four quadrants of the galaxy independently.

Our stellar orbit models require an estimate of the point spread function
(PSF) for each instrument at the time of spectroscopic observations.  
We estimate the PSF for our GMOS data for both NGC 3842 and NGC 4889 from
wide-field images taken during target acquisition.
Although the models discussed herein assume a 0.4-arcsecond PSF for
GMOS data, we have run a small number of models with a 0.7-arcseond PSF and
have found no significant changes in our results.  While observing
NGC 3842, we switched from the OSIRIS spectrograph to the OSIRIS imaging
camera every few hours, and observed the adaptive optics tip/tilt star. 
We can tolerate a large degree of uncertainty in our
measured PSF, as we have re-binned OSIRIS data of NGC 3842 to spatial
scales that are several times coarser than the diffraction-limited FWHM.

\section*{Photometric Data\\}

Our stellar orbit models are constrained to reproduce the observed stellar light profile
of each galaxy, which
requires accurate measurements of each galaxy's surface brightness profile over a
large radial range. 
For radii out to 10
arcseconds, we adopt high-resolution $I$-band (800 nm) surface brightness profiles,
obtained with WFPC2 on the Hubble Space Telescope, and deconvolved with the
instrumental PSF \cite{laine03}. At larger radii out to 115
arcseconds, we use $R$-band (600 nm) data obtained with the 2.1 m telescope at Kitt
Peak National Observatory (KPNO). The KPNO data have a field of view of $5.2 \times 5.2$
arcminutes, which enables accurate sky subtraction. We combine the
individual profiles from WFPC2 and KPNO data at overlapping radii between 5
and 10 arcseconds, accounting for the average $R-I$ color over these
radii. 
To compute the luminosity density profile of each galaxy, we deproject the
surface brightness profile while assuming spheroidal isodensity contours \cite{gebhardt96}. 


\begin{figure}[b!]
\begin{center}
\includegraphics[width=3.8in]{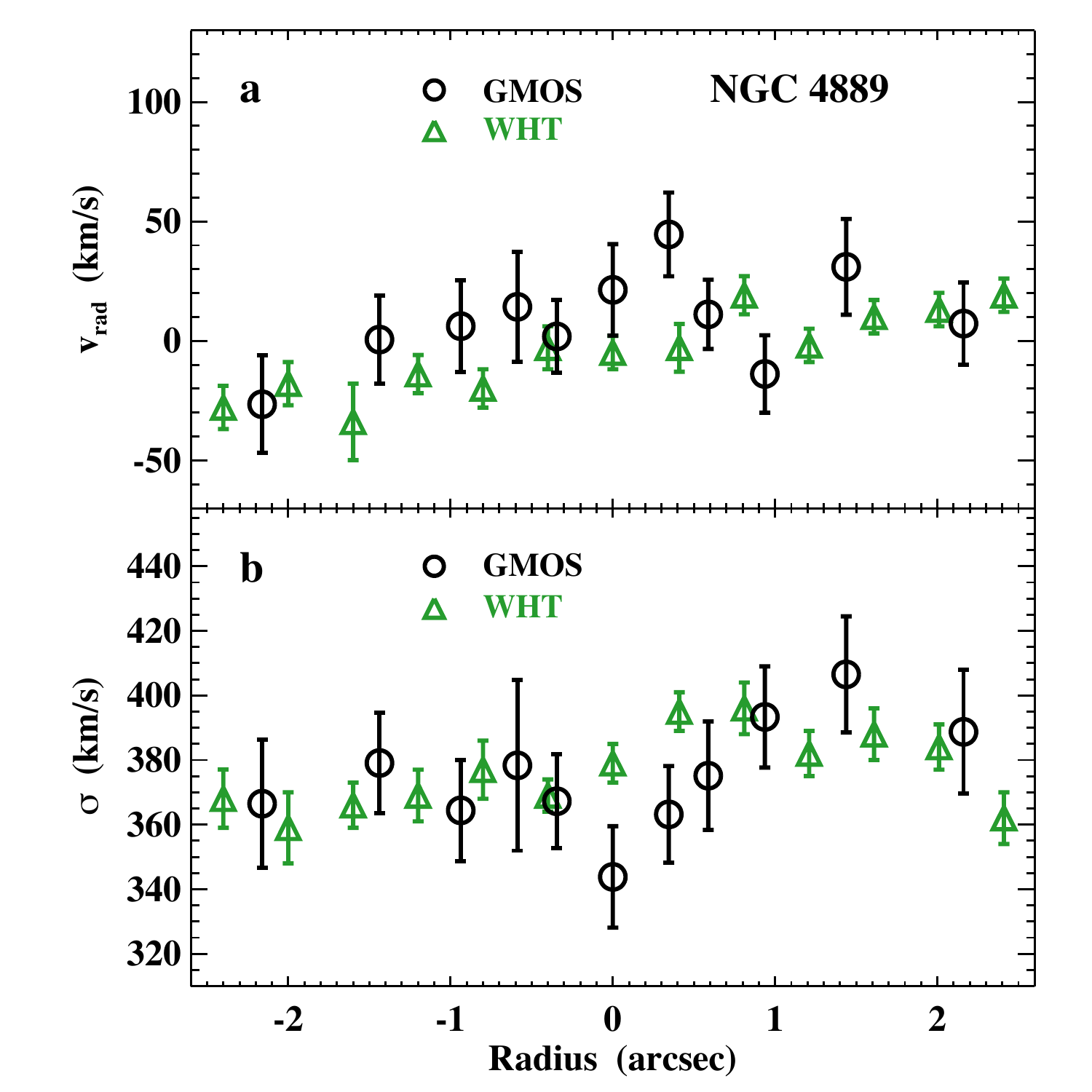}
\end{center}
\vspace{-0.7cm}
\caption{\textbf{(a)} Radial velocity vs. radius in NGC 4889.  \textbf{(b)} Velocity dispersion vs. radius.  Green triangles are measurements from Loubser et al. (2008) \cite{loubser08}, using the ISIS spectrograph on the William Herschel Telescope (WHT).  Black circles are our measurements using the GMOS intrgral-field unit.  Error bars represent one standard deviation.  On the west side and at radii from 1.0 to 2.4 arcseconds on the east side, the GMOS and WHT measurements agree within errors.   At $r < 1.0$ arcseconds on the east side, only GMOS detects a local minimum in velocity dispersion.  The central feature might be unresolved in the WHT data due to worse seeing (1.0 arcseconds).}
\label{fig:GL4889}
\end{figure}

\section*{Kinematic Data from Different Instruments\\}

Stellar kinematics along the major axes of NGC 3842 and NGC 4889 were previously measured by Loubser et al. (2008) \cite{loubser08}, using the ISIS long-slit spectrograph on the William Herschel Telescope (WHT).  The WHT measurements of NGC 3842 are in good agreement with our GMOS measurements for $r \leq 3.0$ arcseconds.  The WHT data for NGC 4889 agree with our GMOS measurements on the west side of NGC 4889 and at radii between 1 and 2.4 arcseconds on the east side.  However, they do not reproduce our measurement of the large central drop in stellar velocity dispersion (Figure~\ref{fig:GL4889}).  For a more direct comparison, we have rebinned our integral-field spectra to match the 0.4-arcsecond spatial sampling and 1.0-arcsecond slit width of the WHT data.  We find that rebinning alleviates the velocity dispersion discrepancy for all but the central point ($r = 0$).  Our detection of a significantly sharper decrease in velocity dispersion is consistent with the superior seeing conditions of our data (0.4 arcseconds for GMOS, versus 1.0 arcseconds for WHT).  Overall, we find our kinematic measurements in NGC 3842 and NGC 4889 to be broadly consistent with the independent measurements by Loubser et al., indicating that our kinematic extraction method has low systematic errors.

For N3842, our VIRUS-P measurements are also consistent with WHT measurements, which extend along the major axis to $r = 20.8$ arcseconds.
%
We prefer using data from VIRUS-P in our stellar orbit models because
they extend to larger radii and provide full two-dimensional spatial
sampling. 

Near the center of NGC 3842, OSIRIS and GMOS provide independent measurements of stellar kinematics.  
We have binned data from OSIRIS and GMOS at identical spatial scales out to $r = 0.7$
arcseconds (330 pc) and have run orbit models fitting LOSVDs from OSIRIS
and GMOS simultaneously (as well as VIRUS-P data at large
radii). Including the OSIRIS data causes the best-fit value of $\mbh$ to
decrease by up to 23\%, and the best-fit value of $\mlstars$ to increase by
as much as 8\%. This occurs because OSIRIS data show a less drastic
increase in velocity dispersion than data from GMOS.  In spite of these
differences, results with and without OSIRIS data are consistent at the
68\% confidence level.

Models fitting OSIRIS and GMOS data together yield higher average $\chi^2$ values
per LOSVD. This is true even if we ignore the central regions where LOSVDs
from OSIRIS and GMOS are not fully consistent. Even with several template
stars, the overlapping absorption features in the H-band spectral region
are difficult to model, and the LOSVDs derived from OSIRIS data may have
systematic errors. Consequently, we judge the models with only GMOS and
VIRUS-P data to be more reliable.

\section*{Stellar Orbit Models and Statistical Analysis\\}

\begin{figure}[b!]
\begin{center}
\includegraphics[width=3.3in]{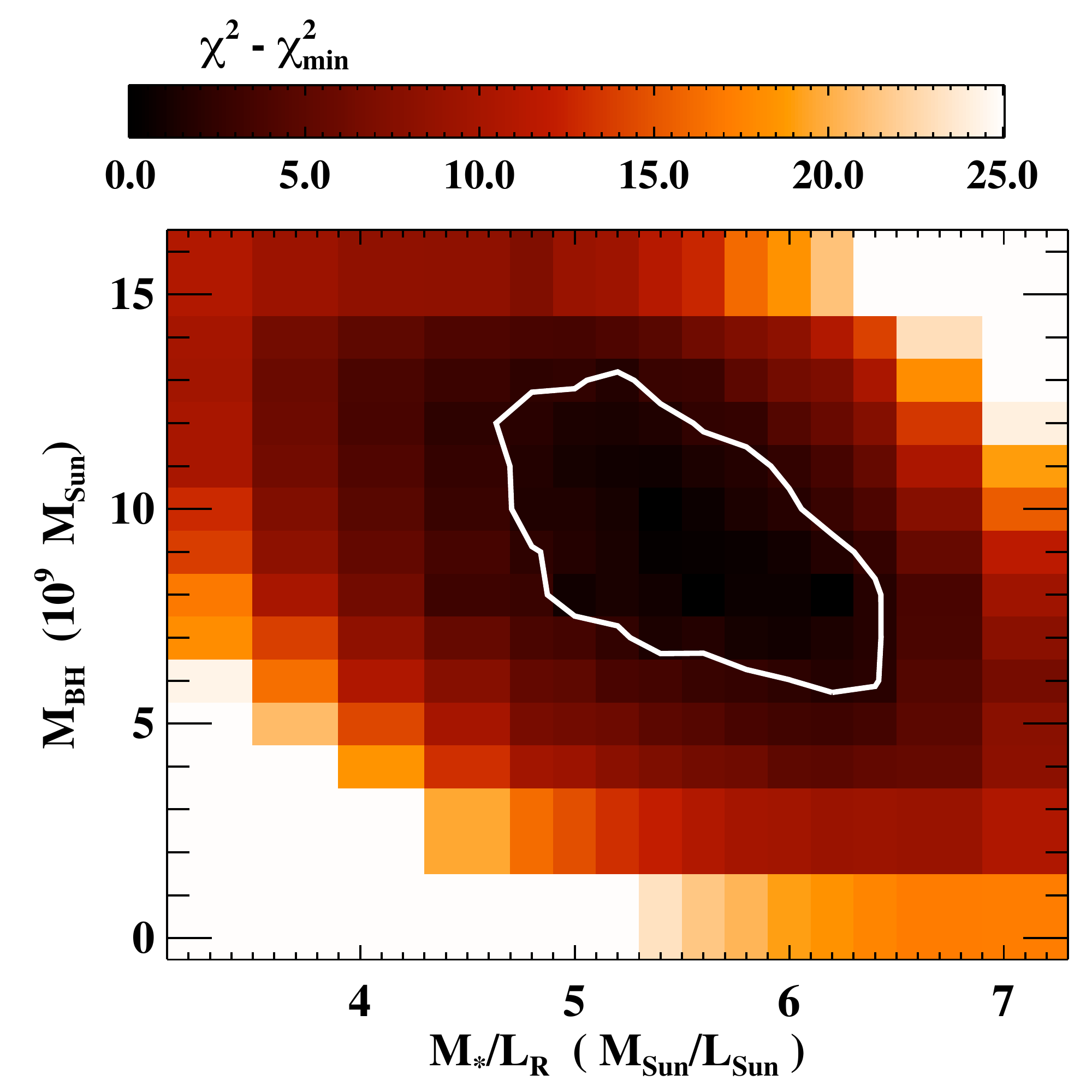}
\end{center}
\vspace{-0.7cm}
\caption{Map of $\chi^2$ versus $\mbh$ and $\mlstars$ for stellar orbit models of NGC 3842.  The models fit data from GMOS and VIRUS-P.
The diagonal trend in $\chi^2$ indicates the degeneracy between stellar mass and black hole mass near the center of NGC 3842.  For two free parameters with Gaussian likelihood distributions, the 68\% confidence interval is defined where $\chi^2 - \chi^2_{min} \leq  2.30$, illustrated by the thick white contour.  We 
obtain 68\% confidence intervals of $7.2 - 12.7 \times 10^9 \msun$ for $\mbh$, and $4.4 - 5.8 \msun/\lsun$ for $\mlstars$.  The median values, which we adopt as our final measurements, are $\mbh = 9.7 \times 10^9 \msun$ and $\mlstars  = 5.1 \msun/\lsun$.}
\label{fig:chi2}
\end{figure}

We generate models of NGC 3842 and NGC 4889 using Schwarzschild's method
\cite{schwarz}, in which test particle orbits are computed in a static
axisymmetric gravitational potential.  We assume that each galaxy contains
three mass components: stars, a central black hole, and an extended dark
matter halo. The stellar mass density is assumed to follow the same profile
as the observed luminosity density,
with a constant stellar mass-to-light ratio, $\mlstars$.  
Our modeling procedures are similar to those for NGC 6086, described in McConnell et
al. (2011) \cite{mcconnell11}.\\
\indent Each orbit in the model is assigned a scalar weight, and the set of
best-fit orbital weights is determined by comparing projected LOSVDs from
the orbits to the observed LOSVDs for the galaxy. Each observed LOSVD
spatially maps to a linear combination of bins within the model, according
to the spatial boundaries of the corresponding spectrum and the PSF of the
observations. A corresponding model LOSVD is computed from the projected
velocity distributions of individual orbits in each spatial bin, the
appropriate combination of spatial bins, and the orbital weights. The
best-fit weights are determined by the method of maximum entropy \cite{RT88}, with the fixed constraint that the summed spatial
distribution of all weighted orbits must match the observed luminosity
density profile.
The essential output of each model is a measurement of $\chi^2$, which defines
the goodness of fit between our observed LOSVDs and the model LOSVDs, using
the optimal combination of orbital weights.  We determine the best-fit values
and confidence intervals in $\mbh$ and $\mlstars$ by evaluating the relative likelihood between models with different assumed values of $\mbh$ and $\mlstars$.   Figure~\ref{fig:chi2} illustrates the behavior of $\chi^2$ with respect to $\mbh$
and $\mlstars$, for models with our best fitting dark matter halo.

\begin{figure}[b!]
\begin{center}
\hspace{-0.7cm}
\includegraphics[width=3.8in]{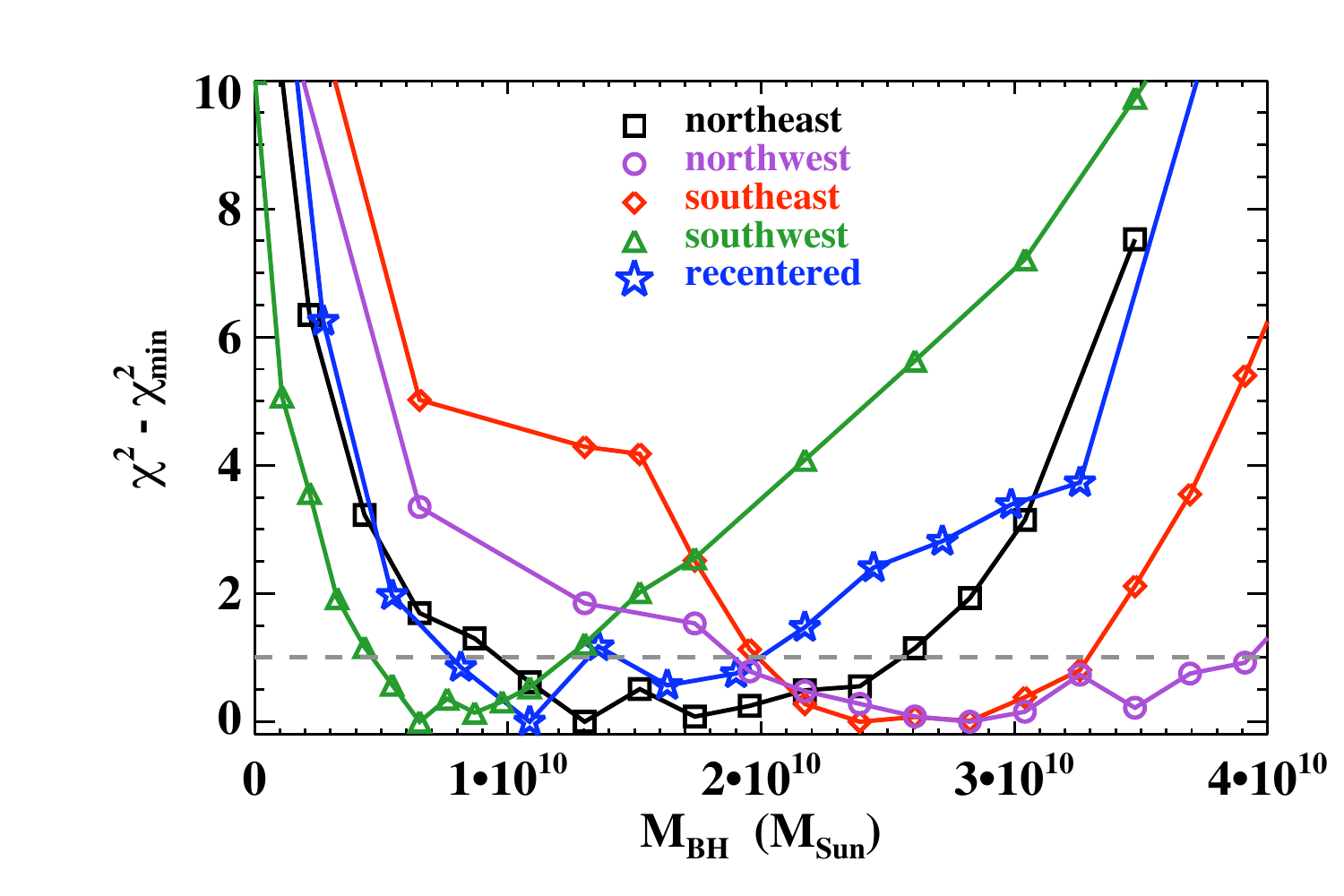}
\end{center}
\vspace{-0.7cm}
\caption{$\chi^2$ vs. $\mbh$ for NGC 4889, after marginalizing over $\mlstars$.  Each line with symbols represents a models constrained by different set of LOSVDs from GMOS.  Black squares, purple circles, red diamonds, and green triangles each use LOSVDs from a different quadrant of the galaxy.  Blue stars represent models with spatially offset LOSVDs, to match the largest velocity dispersion with the center of the gravitational potential.  Considering all models, the 68\% confidence interval for $\mbh$ is $0.6 - 3.7 \times 10^{10} \msun$.}
\label{fig:chi4889}
\end{figure}

\section*{Models of NGC 4889\\}

Stellar kinematics in NGC 4889 are asymmetric with respect to the major and minor axes of
the galaxy.  Integral-field data from GMOS reveals velocity dispersions above $410 \kms$ on the east side of
the galaxy, while the velocity dispersion rarely exceeds $380 \kms$ on the
west side.  This asymmetry prevents NGC 4889 from being
fully described by a single set of axisymmetric orbit models.  In order to
place upper and lower bounds on the central black hole mass, we have run
four suites of models, each fitting kinematics from one projected quadrant of NGC 4889.  The northeast, southeast, and northwest quadrants yield consistent black hole masses, spanning a 68\% confidence interval of $\mbh = 1.0 - 3.7 \times 10^{10} \msun$.  The southwest quadrant has a maximum velocity dispersion of 373 km/s and yields a 68\% confidence interval of $\mbh = 0.6 - 1.7 \times 10^{10} \msun$.  

We have run an additional set of models to approximate $\mbh$ in the case of an
off-center black hole.  We apply a constant spatial offset of 1.4
arcseconds (700 pc) to the kinematics on the east side of the galaxy, such
that the highest velocity dispersion is aligned with the center of the
model gravitational potential.  These models cannot be fully trusted
because the kinematic and photometric data are misaligned.  Still, the
resulting 68\% confidence interval for $\mbh$ falls entirely within the
range bracketed by the models from different quadrants.
Results from individual trials are listed in Table~\ref{tab:n4889}.  
Figure~\ref{fig:chi4889} illustrates $\chi^2$ versus $\mbh$ for each series
of models, after marginalizing over $\mlstars$.

\begin{figure}[b!]
\begin{center}
\includegraphics[width=3.5in]{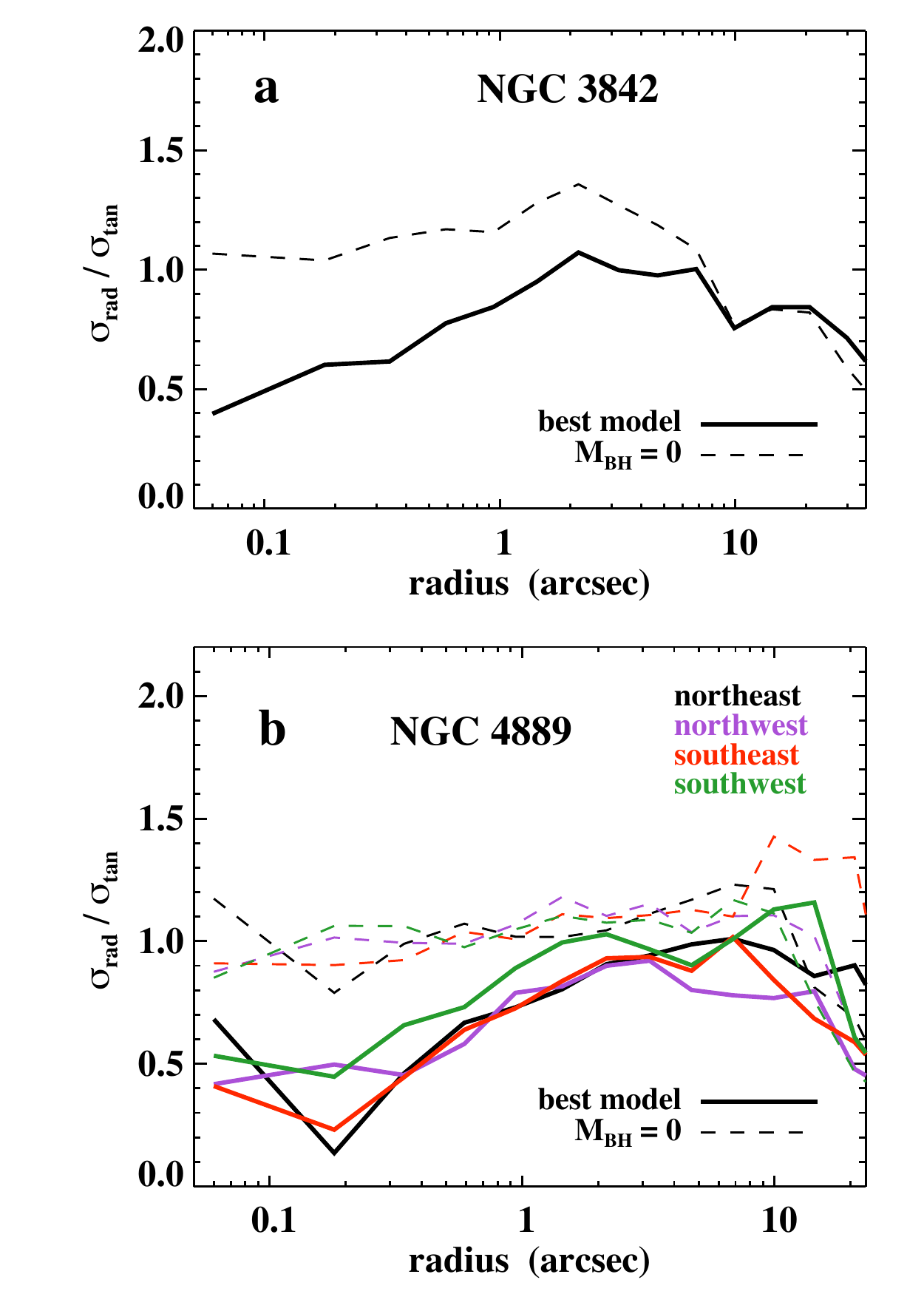}
\end{center}
\vspace{-0.7cm}
\caption{Orbital anisotropy in models of NGC 3842 and NGC 4889.  
Three-dimensional stellar velocities in the models are divided into radial and tangential components.  The ratio of velocity dispersions, $\sigma_{\rm rad} / \sigma_{\rm tan}$, varies with radius and the assumed $\mbh$ in the model.  
\textbf{(a)}  The thick solid line represents the best-fitting model for NGC 3842, with $\mbh =  8.5 \times 10^9 \msun$.  The thin dashed line represents the best model with no black hole.  \textbf{(b)}  Thick solid lines represent the best-fitting model for each quadrant in NGC 4889, with $\mbh = 1.3 \times 10^{10} \msun$ (northeast), $2.8 \times 10^{10} \msun$ (northwest), $2.4 \times 10^{10} \msun$ (southeast), and $6.5 \times 10^9 \msun$ (southwest).  Thin dashed lines represent models with no black hole.  For NGC 3842 and each quadrant of NGC 4889, our kinematic data are best fit with a massive black hole and a prevalence of tangential orbits at $r < 2$ arcseconds (1 kpc).  In NGC 4889, this tangential bias is responsible for the observed decrease in line-of-sight velocity dispersion near the center.}
\label{fig:bias}
\end{figure}

Although three-dimensional stellar velocities must increase in the vicinity of a black hole, a deficiency of radial orbits can produce a central minimum in the line-of-sight velocity dispersion, as we observe in NGC 4889.  Indeed, our best-fitting models of NGC 4889 exhibit tangential bias at small radii, as shown in Figure~\ref{fig:bias}.  In contrast, models without a black hole reproduce the central drop in velocity dispersion with a nearly isotropic orbital distribution.  However, these models yield a worse overall fit, indicated by higher values of $\chi^2$ in each quadrant.  Models of NGC 3842 exhibit a  similar but less severe trend, consistent with the modest increase in line-of-sight velocity dispersion toward the center.


The individual quadrants of NGC 4889 represent large variations in stellar kinematics, but each quadrant still partially constrains the enclosed mass within the central few arcseconds.  By adopting the most extreme range of confidence limits, $\mbh = 0.6 - 3.7 \times 10^{10} \msun$, we only exclude black holes whose gravitational influence would contradict our entire field of data.  Further extensions to this confidence interval should only reflect overall systematic biases.  Large systematic biases in our kinematic measurements are unlikely, as demonstrated by their agreement with independent measurements by Loubser et al. (2008) \cite{loubser08}.  
Our models of NGC 4889 assume an edge-on inclination.  This is indirectly supported by the observed axis ratio of 0.7, which implies a relatively eccentric intrinsic shape even for an edge-on system.  Models with a more face-on inclination might yield a systematically higher black hole mass \cite{vdb10}.  A fundamental assumption of all orbit superposition models is that the stellar motions reflect a steady-state gravitational potential, rather than transient conditions.  
These models could misrepresent the range of allowed black hole masses if the observed kinematics in NGC 4889 reflected a temporary phenomenon such as an ongoing galaxy merger.  
NGC 4889 appears photometrically undisturbed, 
reducing the likelihood of such an event.\\
\indent Because our adopted confidence interval places large error bars on $\mbh$ in NGC 4889, this galaxy has relatively little weight in our fits to the $\mbh-\sigma$ and $\mbh-L$ relationships.  A systematic error in our measurement would produce a minimal bias in the best-fit relations.  Likewise, our discussion of steepening and scatter at the high-mass ends of the correlations depends upon several objects and is not highly sensitive to the measurement in NGC 4889.

\section*{Power-Law Fits to the $\mbh-\sigma$ and $\mbh-L$ Relations\\}

We revisit the $\mbh-\sigma$ and $\mbh-L$ relations by updating the sample
of 49 black holes from G\"{u}ltekin et al. (2009; hereafter G09)
\cite{gultekin09}, which was compiled from earlier studies. Including NGC
3842 and NGC 4889, we add 17 galaxies with recently
measured black hole masses to the sample.  The new objects include two more 
BCGs \cite{nowak08,mcconnell11}, eight active galactic nuclei with high-precision
maser-based measurements \cite{greene10, kuo11}, and two galaxies with
pseudobulges \cite{kormendy11}. We also include updated black hole masses
for 16 other galaxies in the 2009 sample, based on stellar orbit
models with dark matter halos and more thorough orbit libraries
\cite{shen10,vdb10,gebhardt11,schulze11}.  In particular, the revised
masses for M60 and M87 are twice as large as the earlier values.\\
\indent Our updated sample uses the same selection criteria as G09; in particular,
only direct dynamical measurements of $\mbh$ are included.  
G09 estimated galaxy distances by assuming a Hubble parameter $H_0  = 70 \kms$ Mpc$^{-1}$, and rescaled their sample of black hole masses accordingly ($\mbh \propto D$).  We have followed this convention for NGC 3842, NGC 4889, and the rest of our sample.  Our fits to
$\mbh(\sigma)$ do not include upper limits.  Updated models of one galaxy,
NGC 2778, do not produce a significant black hole detection; after removing
this object, our updated sample contains 65 black hole masses.\\
\indent We define $\sigma$ in the same manner as the 2009 sample.  Wherever
possible, we use the luminosity-weighted effective velocity dispersion,
measured using spatially resolved data out to one effective radius. G09
found no evidence of systematic bias between this definition of $\sigma$
and more ubiquitous single-aperture measurements.  Nonetheless, several
galaxies in G09 have measurements of $\sigma$ that include data at very
small radii, within which the central black hole directly influences the
stellar velocity dispersion.  This is inappropriate for studies that wish
to treat $\mbh$ and $\sigma$ as fundamentally independent variables.  We
have therefore re-evaluated $\sigma$ in three galaxies with large black hole
masses and available spatially resolved kinematics, by excluding data
within the black hole's radius of influence.  We use this same treatment to
measure $\sigma$ for NGC 3842 and NGC 4889.  For M87, we adopt the updated values of $\mbh$ and $\sigma$ from Gebhardt et al. (2011) \cite{gebhardt11}.  Following G09, we assume all
measurements of $\sigma$ have an uncertainty of at least 5\%.\\
\indent To fit the $\mbh-L$ relationship, we only consider early-type galaxies for
which the stellar luminosity of the spheroidal component can be measured
reliably.  Including NGC 3842, we add 6 galaxies to the $\mbh-L$ sample of
G09.  We exclude NGC 2778.   We also
exclude NGC 3607 and NGC 4564, for which the literature contains large
discrepancies in the measured luminosity \cite{lauer07, gultekin09}.  Our
final sample for fitting $\mbh-L$ contains 36 black hole masses.\\
\indent Our fits to $\mbh-\sigma$ and $\mbh-L$ assume a single-index power law as
the functional form of both relations, following the convention most
commonly used in prior studies.  Specifically, we define the $\mbh-\sigma$
relationship to be $\log_{10}(\mbh/\msun) = \alpha + \beta \log_{10}(\sigma
/ 200 \kms)$, and the $\mbh - L$ relationship to be $\log_{10}(\mbh/\msun)
= \alpha + \beta \log_{10}(L_V / 10^{11} \lsun)$.  For each relationship,
we follow the method of Tremaine et al. (2002) \cite{tremaine02} to fit for
$\alpha$ and $\beta$.  We minimize the quantity
\begin{equation}
\chi^2 = \sum_{i=1}^{N} \frac{\left(  \mbh ,_i - \alpha - \beta \sigma_i \right) ^2 }{\epsilon_0^2 + \epsilon_{M,i}^2 + \beta^2 \epsilon_{\sigma,i}^2}
\end{equation}
where $\epsilon_\sigma$ is the measurement error in $\sigma$, $\epsilon_M$
is the measurement error in $\mbh$, and $\epsilon_0$ is the intrinsic
scatter in the $\mbh - \sigma$ relation. We set $\epsilon_0$ such that
$\chi^2$ per degree of freedom is unity after minimization.  The 68\%
confidence intervals for $\alpha$ and $\beta$ correspond to the maximum
range of $\alpha$ and $\beta$ for which $\chi^2 - \chi^2_{min} \le 1$.  

We list our best-fit values of $\alpha$, $\beta$, and $\epsilon_0$ for various sub-samples in Table~\ref{tab:fits}.  We list our full sample of galaxies in Table~\ref{tab:gals}.  Graham et al. (2011) \cite{graham11} recently compiled a sample of 64 galaxies with dynamical black hole measurements and found a power-law index of 5.13 for the $\mbh - \sigma$ relation, very similar to the index we report herein.  However, only 52 galaxies appear in their sample as well as ours.  Of the 13 galaxies that appear only in our sample, seven use precise maser-based measurements of $\mbh$ \cite{greene10, kuo11}, and three are BCGs.  Additionally, 11 galaxies have received updated measurements of $\mbh$ since the compilation of Graham et al. (2011), using stellar orbit models with dark matter halos and larger orbit libraries \cite{schulze11}.





\clearpage

\begin{table*}[htbp]
\begin{center}
\caption{Spectroscopic observations}
\label{tab:spec}
\begin{tabular}[b]{cccccccccc}  
\\
\hline
Galaxy & Instrument & UT Date & $N_{\rm exp}$ & $t_{\rm exp}$ & $\lambda$ & $\Delta\lambda$  & $r_{\rm max}$ & $\Delta r$ & PSF FWHM\\
&  &  &  & (s) & (nm) & (nm) & (arcsec) & (arcsec) & (arc sec)\\
\hline 
\\
NGC 3842 & GMOS & April 27, 2003 & 5 & 1200 & 755-949 & 0.25 & 3.8 & 0.2 & 0.4\\
NGC 3842 & OSIRIS & May 8-10, 2010 & 16 & 900 & 1473-1803 & 0.72 & 0.7 & 0.05 & 0.08 \\
NGC 3842 & VIRUS-P & March 7-8, 2011& 3 & 1200 & 358-589 & 0.52 & 35.3 & 4.1 & 2.0 \\
&&&&&&&&&
\\
NGC 4889 & GMOS & March 13, 2003 & 5 & 1200 & 755-949 & 0.25 & 3.8 & 0.2 & 0.4\\
\hline
\end{tabular}
\end{center}
\textbf{Notes:}  $N_{\rm exp}$ is the number of science exposures recorded.  $t_{\rm exp}$ is the integration time per exposure.  $\Delta\lambda$ is the median instrumental resolution (FWHM).  $r_{\rm max}$ is the maximum radius of usable data, with respect to the center of the galaxy.  $\Delta r$ is the instrumental spatial scale, equal to the angular diameter of one lenslet or fiber.
\end{table*}

\begin{table*}[htbp]
\begin{center}
\caption{Models of NGC 4889}
\label{tab:n4889}
\begin{tabular}[b]{ccccccc}  
\\
\hline
Quadrant & $\mbh$ & $\mbh ,_{\rm min}$ & $\mbh ,_{\rm max}$ & $\mlstars$ & $\mlstars ,_{\rm min}$ & $\mlstars ,_{\rm max}$ \\
& ($\msun$) & ($\msun$) & ($\msun$) & ($\msun/\lsun$) & ($\msun/\lsun$) & ($\msun/\lsun$) \\
\hline 
\\
northeast & $1.7 \times 10^{10}$ & $1.0 \times 10^{10}$ & $2.5 \times 10^{10}$ & $6.1$ & 4.6 & 7.3 \\
southeast & $2.6 \times 10^{10}$ & $2.0 \times 10^{10}$ & $3.2 \times 10^{10}$ & $5.6$ & 4.2 & 6.7 \\
northwest & $2.7 \times 10^{10}$ & $1.6 \times 10^{10}$ & $3.7 \times 10^{10}$ & $5.8$ & 4.4 & 7.0 \\
southwest & $9.8 \times 10^9$ & $5.5 \times 10^9$ & $1.7 \times 10^{10}$ & $6.6$ & 5.3 & 7.6 \\
\\
east & $2.9 \times 10^{10}$ & $2.1 \times 10^{10}$ & $3.4 \times 10^{10}$ & $5.4$ & 4.5 & 6.4\\
west & $1.2 \times 10^{10}$ & $6.5 \times 10^9$ & $2.0 \times 10^{10}$ & $6.4$ & 5.2 & 7.4\\
\\
recentered & $1.5 \times 10^{10}$ & $8.7 \times 10^9$ & $2.4 \times 10^{10}$ & $6.5$ & 5.4 & 7.3 \\
\hline
\end{tabular}
\end{center}
\textbf{Notes:}  The ``east'' and ``west'' trials used LOSVDs from spectra that were binned symmetrically over the north and south sides of the galaxy.  The ``recentered'' trial added an artificial position offset to the LOSVDs, such that the maximum velocity dispersion was placed at the center of the gravitational potential.  $\mbh ,_{\rm min}$, $\mbh ,_{\rm max}$, $\mlstars ,_{\rm min}$, and $\mlstars ,_{\rm max}$ represent $68\%$ confidence limits.  $\mlstars$ corresponds to $R$-band photometry.
\end{table*}

\begin{table*}[htbp]
\begin{center}
\caption{Fits to $\mbh - \sigma$ and $\mbh - L$}
\label{tab:fits}
\begin{tabular}[b]{cccccc}  
\\
\hline
Relationship & Sample & $N_{\rm gal}$ & $\alpha$ & $\beta$ & $\epsilon_0$ \\
\hline 
\\
$\mbh-\sigma$ & all & 65 & $8.29 \pm 0.06$ & $5.12 \pm 0.36$  & 0.43 \\
$\mbh-\sigma$ & early-type & 45 & $8.38 \pm 0.06$ & $4.53 \pm 0.40$ & 0.38 \\
$\mbh-\sigma$ & late-type & 20 & $7.97 \pm 0.22$ & $4.58 \pm 1.25$ & 0.44 \\
$\mbh-\sigma$ & ML & 36 & $8.43 \pm 0.07$ & $4.66 \pm 0.43$ & 0.38 \\
&&&&&
\\
$\mbh-L$ & ML & 36 & $9.16 \pm 0.11$ & $1.16 \pm 0.14$ & 0.50 \\
\hline
\end{tabular}
\end{center}
\textbf{Notes:}  $N_{\rm gal}$ is the number of galaxies in the sample.  ML refers to the sample of early-type galaxies with reliable spheroid luminosity measurements.
\end{table*}

\begin{table*}
\begin{center}
\caption{Galaxies with dynamical measurements of $\mbh$}
\label{tab:gals}

\begin{tabular}[b]{lccccccccc}  
\\
\hline
\\
Galaxy & Type & $D$ & $\sigma$ & log$_{10}$($L_{V,\rm sph}$) & $\mbh$ & $\mbh ,_{\rm min}$ & $\mbh ,_{\rm max}$ & Method & Ref. \\
\\
& & (Mpc) & ($\kms$) & log$_{10}$($\lsun$) & ($\msun$) & ($\msun$) & ($\msun$) & & \\
\hline 
Milky Way $^a$ & Sbc & 0.008 & $103 \pm 20$ & & $4.1 \times 10^6$ & $3.5 \times 10^6$ & $4.7 \times 10^6$ & stars & \cite{gultekin09,ghez08,gillessen09} \\
A1836-BCG & E & 157.5 & $288 \pm 14$ & 11.26 & $3.9 \times 10^9$ & $3.3 \times 10^9$ & $4.3 \times 10^9$ & gas & \cite{gultekin09,dallabonta09} \\
A3565-BCG $^b$ & E & 54.4 & $322 \pm 16$ & 11.24 & $1.4 \times 10^9$ & $1.2 \times 10^9$ & $1.7 \times 10^9$ & gas & \cite{dallabonta09} \\
Circinus & Sb & 4.0 & $158 \pm 18$ & & $1.7 \times 10^6$ & $1.4 \times 10^6$ & $2.1 \times 10^6$ & masers & \cite{gultekin09,greenhill03} \\
IC1459 $^c$ & E4 & 30.9 & $315 \pm 16$ & 10.96 & $2.8 \times 10^9$ & $1.6 \times 10^9$ & $3.9 \times 10^9$ & stars & \cite{gultekin09,cappellari02} \\
N221 (M32) & E2 & 0.86 & $75 \pm 3$ & 8.66 & $3.1 \times 10^6$ & $2.5 \times 10^6$ & $3.7 \times 10^6$ & stars & \cite{gultekin09,verolme02} \\
N224 (M31) & Sb & 0.80 & $160 \pm 8$ & & $1.5 \times 10^8$ & $1.2 \times 10^8$ & $2.4 \times 10^8$ & stars & \cite{gultekin09,bender05} \\
N524 & S0 & 23.3 & $235 \pm 12$ & 10.67 & $8.3 \times 10^8$ & $7.9 \times 10^8$ & $9.2 \times 10^8$ & stars & \cite{krajnovic09} \\
N821 & E4 & 25.5 & $209 \pm 10$ & 10.43 & $1.8 \times 10^8$ & $1.0 \times 10^8$ & $2.6 \times 10^8$ & stars & \cite{schulze11} \\
N1023 & SB0 & 12.1 & $205 \pm 10$ & 10.18 & $14.6 \times 10^7$ & $4.1 \times 10^7$ & $5.1 \times 10^7$ & stars & \cite{gultekin09,bower01} \\
N1068 (M77) & SB & 15.4 & $151 \pm 7$ & & $8.6 \times 10^6$ & $8.3 \times 10^6$ & $8.9 \times 10^6$ & masers & \cite{gultekin09,LB03} \\
N1194 $^{b,d}$ & SA0+ & 55.5 & $148^{+26}_{-22}$ & & $6.8 \times 10^7$ & $6.5 \times 10^7$ & $7.1 \times 10^7$ & masers & \cite{kuo11} \\
N1300 & SB(rs)bc & 20.1 & $218 \pm 10$ & & $7.1 \times 10^7$ & $3.6 \times 10^7$ & $1.4 \times 10^8$ & gas & \cite{gultekin09,atkinson05} \\
N1316 & E & 18.6 & $226 \pm 11$ & 11.26 & $1.5 \times 10^8$ & $1.24 \times 10^8$ & $1.75 \times 10^8$ & stars & \cite{nowak08} \\
N1332 & S0 & 22.3 & $328 \pm 16$ & 10.14 & $1.45 \times 10^9$ & $1.25 \times 10^9$ & $1.65 \times 10^9$ & stars & \cite{rusli11} \\
N1399 $^{e,f}$ & E1 & 21.1 & $296 \pm 15$ & 10.78 & $5.1 \times 10^8$ & $4.4 \times 10^8$ & $5.8 \times 10^8$ & stars & \cite{gultekin09,gebhardt07} \\
N1399 $^{e,f}$ & E1 & 21.1 & $296 \pm 15$ & 10.78 & $1.3 \times 10^9$ & $6.4 \times 10^8$ & $1.8 \times 10^9$ & stars & \cite{gultekin09,houghton06} \\
N2273 $^{b,d}$ & SB(r)a & 26.8 & $144^{+18}_{-16}$ & & $7.8 \times 10^6$ & $7.4 \times 10^6$ & $8.2 \times 10^6$ & masers & \cite{kuo11} \\
N2549 & S0 & 12.3 & $145 \pm 7$ & 9.60 & $1.4 \times 10^7$ & $1.0 \times 10^7$ & $1.47 \times 10^7$ & stars & \cite{krajnovic09} \\
N2748 & Sc & 24.9 & $115 \pm 5$ & & $4.7 \times 10^7$ & $8.6 \times 10^6$ & $8.5 \times 10^7$ & gas & \cite{gultekin09,atkinson05} \\
N2787 & SB0 & 7.9 & $189 \pm 9$ & & $4.3 \times 10^7$ & $3.8 \times 10^7$ & $4.7 \times 10^7$ & gas & \cite{gultekin09,sarzi01} \\
N2960 $^{b,d}$ & Sa & 75.3 & $166^{+16}_{-15}$ & & $1.21 \times 10^7$ & $1.16 \times 10^7$ & $1.26 \times 10^7$ & masers & \cite{kuo11} \\
N3031 (M81) & Sb & 4.1 & $143 \pm 7$ & & $8.0 \times 10^7$ & $6.9 \times 10^7$ & $1.0 \times 10^8$ & gas & \cite{gultekin09,devereux03} \\
N3115 & S0 & 10.2 & $230 \pm 11$ & 10.40 & $9.6 \times 10^8$ & $6.7 \times 10^8$ & $1.5 \times 10^9$ & stars & \cite{gultekin09,emsellem99} \\
N3227 & SBa & 17.0 & $133 \pm 12$ & & $1.5 \times 10^7$ & $7.0 \times 10^6$ & $2.0 \times 10^7$ & stars & \cite{gultekin09,davies06} \\
N3245 & S0 & 22.1 & $205 \pm 10$ & & $2.2 \times 10^8$ & $1.7 \times 10^8$ & $2.7 \times 10^8$ & gas & \cite{gultekin09,barth01} \\
N3368 $^g$ & SAB(rs)ab & 10.4 & $122^{+28}_{-24}$ &  & $7.5 \times 10^6$ & $6.0 \times 10^6$ & $9.0 \times 10^6$ & stars & \cite{nowak10} \\
N3377 & E6 & 11.7 & $145 \pm 7$ & 9.98 & $1.9 \times 10^8$ & $9.0 \times 10^7$ & $2.9 \times 10^8$ & stars & \cite{schulze11} \\
N3379 (M105) $^b$ & E0 & 11.7 & $206 \pm 10$ & 10.37 & $4.6 \times 10^8$ & $3.4 \times 10^8$ & $5.7 \times 10^8$ & stars & \cite{vdb10} \\
N3384 & E1 & 11.7 & $143 \pm 7$ & 9.90 & $1.1 \times 10^7$ & $6.0 \times 10^6$ & $1.6 \times 10^7$ & stars & \cite{schulze11} \\
N3393 $^b$ & SB(rs) & 53.6 & $148 \pm 10$ & & $3.3 \times 10^7$ & $3.1 \times 10^7$ & $3.5 \times 10^7$ & masers & \cite{kondratko08} \\
N3489 $^g$ & SAB(rs)0+ & 12.1 & $100^{+15}_{-11}$ &  & $6.0 \times 10^6$ & $5.2 \times 10^6$ & $6.8 \times 10^6$ & stars & \cite{nowak10} \\
N3585 & S0 & 21.2 & $213 \pm 10$ & 10.65 & $3.4 \times 10^8$ & $2.8 \times 10^8$ & $4.9 \times 10^8$ & stars & \cite{gultekin09,gultekin09b} \\
N3607 $^h$ & E1 & 19.9 & $229 \pm 11$ & & $1.2 \times 10^8$ & $7.9 \times 10^7$ & $1.6 \times 10^8$ & stars & \cite{gultekin09,gultekin09b} \\
N3608 & E1 & 23.0 & $182 \pm 9$ & 10.35 & $4.7 \times 10^8$ & $3.7 \times 10^8$ & $5.7 \times 10^8$ & stars & \cite{schulze11} \\
N3842 $^i$ & E & 98.4 & $270 \pm 14$ & 11.20 & $9.7 \times 10^9$ & $7.2 \times 10^9$ & $1.27 \times 10^{10}$ & stars & \\
N3998 & S0 & 14.9 & $305 \pm 15$ & & $2.4 \times 10^8$ & $6.2 \times 10^7$ & $4.5 \times 10^8$ & gas & \cite{gultekin09,defrancesco06} \\
N4026 & S0 & 15.6 & $180 \pm 9$ & 9.86 & $2.1 \times 10^8$ & $1.7 \times 10^8$ & $2.8 \times 10^8$ & stars & \cite{gultekin09,gultekin09b} \\
N4258 & SABbc & 7.2 & $115 \pm 10$ & & $3.78 \times 10^7$ & $3.77 \times 10^7$ & $3.79 \times 10^7$ & masers & \cite{gultekin09,herrnstein05} \\
N4261 & E2 & 33.4 & $315 \pm 15$ & 11.02 & $5.5 \times 10^8$ & $4.3 \times 10^8$ & $6.6 \times 10^8$ & gas & \cite{gultekin09,ferrarese96} \\
N4291 & E2 & 25.0 & $242 \pm 12$ & 10.20 & $9.2 \times 10^8$ & $6.3 \times 10^8$ & $1.21 \times 10^9$ & stars & \cite{schulze11} \\
N4342 & S0 & 18.0 & $225 \pm 11$ &  & $3.6 \times 10^8$ & $2.4 \times 10^8$ & $5.6 \times 10^8$ & stars & \cite{gultekin09,cretton99} \\
N4374 (M84) & E1 & 17.0 & $296 \pm 14$ & 10.91 & $8.5 \times 10^8$ & $7.7 \times 10^8$ & $9.4 \times 10^8$ & gas & \cite{walsh10} \\
N4388 $^{b,d}$ & SA(s)b & 19.8 & $107^{+8}_{-7}$ & & $8.8 \times 10^6$ & $8.6 \times 10^6$ & $9.0 \times 10^6$ & masers & \cite{kuo11} \\
N4459 & E2 & 17.0 & $167 \pm 8$ & 10.36 & $7.4 \times 10^7$ & $6.0 \times 10^7$ & $8.8 \times 10^7$ & gas & \cite{gultekin09,sarzi01} \\
N4473 & E4 & 17.0 & $190 \pm 9$ & 10.39 & $1.0 \times 10^8$ & $5.0 \times 10^7$ & $1.5 \times 10^8$ & stars & \cite{schulze11} \\
N4486 (M87) $^{b,j}$ & E1 & 17.0 & $324^{+28}_{-16}$ & 11.10 & $6.3 \times 10^9$ & $5.9 \times 10^9$ & $6.6 \times 10^9$ & stars & \cite{gebhardt11} \\
N4486A & E2 & 17.0 & $111 \pm 5$ & 9.41 & $1.3 \times 10^7$ & $9.0 \times 10^6$ & $1.8 \times 10^7$ & stars & \cite{gultekin09,nowak07} \\
N4564 $^k$ & S0 & 17.0 & $162 \pm 8$ & & $9.4 \times 10^7$ & $6.8 \times 10^7$ & $1.2 \times 10^8$ & stars & \cite{schulze11} \\
N4594 (M104) & Sa & 10.3 & $240 \pm 12$ & & $5.3 \times 10^8$ & $4.74 \times 10^8$ & $6.08 \times 10^8$ & stars & \cite{kormendy11} \\
\\
\hline
\end{tabular}

Continued on next page.

\end{center}
\end{table*}

\begin{table*}
\begin{center}

\textbf{Table 4, continued}\\


\begin{tabular}[b]{lccccccccc}  
\\
\hline
Galaxy & Type & $D$ & $\sigma$ & log$_{10}$($L_{V,\rm sph}$) & $\mbh$ & $\mbh ,_{\rm min}$ & $\mbh ,_{\rm max}$ & Method & Ref. \\
\\
& & (Mpc) & ($\kms$) & log$_{10}$($\lsun$) & ($\msun$) & ($\msun$) & ($\msun$) & & \\
\hline 
\\
N4596 & SB0 & 18.0 & $136 \pm 6$ & & $8.4 \times 10^7$ & $5.9 \times 10^7$ & $1.2 \times 10^8$ & gas & \cite{gultekin09,sarzi01} \\
N4649 (M60) $^{b,l}$ & E2 & 16.5 & $341 \pm 17$ & 10.99 & $4.7 \times 10^9$ & $3.7 \times 10^9$ & $5.8 \times 10^9$ & stars & \cite{shen10} \\
N4697 & E6 & 12.4 & $177 \pm 8$ & 10.45 & $2.0 \times 10^8$ & $1.8 \times 10^8$ & $2.2 \times 10^8$ & stars & \cite{schulze11} \\
N4736 (M94) & Sab & 4.9 & $112 \pm 6$ & & $6.68 \times 10^6$ & $5.14 \times 10^6$ & $8.22 \times 10^6$ & stars & \cite{kormendy11} \\
N4826 (M64) & Sab & 6.4 & $96 \pm 5$ & & $1.36 \times 10^6$ & $1.02 \times 10^6$ & $1.71 \times 10^6$ & stars & \cite{kormendy11} \\
N4889 $^m$ & E & 103.2 & $347 \pm 17$ & 11.42 & $2.1 \times 10^{10}$ & $5.5 \times 10^9$ & $3.7 \times 10^{10}$ & stars &  \\
N5077 & E3 & 44.9 & $222 \pm 11$ & 10.75 & $8.0 \times 10^8$ & $4.7 \times 10^8$ & $1.3 \times 10^9$ & gas & \cite{gultekin09,defrancesco08} \\
N5128 $^e$ & S0/E & 4.4 & $150 \pm 7$ & 10.66 & $3.0 \times 10^8$ & $2.8 \times 10^8$ & $3.4 \times 10^8$ & stars & \cite{gultekin09,silge05} \\
N5128 $^e$ & S0/E & 4.4 & $150 \pm 7$ & 10.66 & $7.0 \times 10^7$ & $3.2 \times 10^7$ & $8.3 \times 10^7$ & stars & \cite{gultekin09,cappellari09} \\
N5576 & E3 & 27.1 & $183 \pm 9$ & 10.44 & $1.8 \times 10^8$ & $1.4 \times 10^8$ & $2.1 \times 10^8$ & stars & \cite{gultekin09,gultekin09b} \\
N5845 & E3 & 28.7 & $234 \pm 11$ & 9.84 & $5.4 \times 10^8$ & $3.7 \times 10^8$ & $7.1 \times 10^8$ & stars & \cite{schulze11} \\
N6086 $^b$ & E & 139.1 & $318 \pm 16$ & 11.18 & $3.8 \times 10^9$ & $2.6 \times 10^9$ & $5.5 \times 10^9$ & stars & \cite{mcconnell11} \\
N6251 & E1 & 106.0 & $290 \pm 14$ & & $6.0 \times 10^8$ & $4.0 \times 10^8$ & $8.0 \times 10^8$ & gas & \cite{gultekin09,ferrarese99} \\
N6264 $^{b,d}$ & S & 145.4 & $158^{+15}_{-14}$ & & $3.03 \times 10^7$ & $2.99 \times 10^7$ & $3.08 \times 10^7$ & masers & \cite{kuo11} \\
N6323 $^{b,d}$ & Sab & 110.5 & $158^{+28}_{-23}$ & & $9.8 \times 10^6$ & $9.7 \times 10^6$ & $9.9 \times 10^6$ & masers & \cite{kuo11} \\
N7052 & E3 & 70.9 & $266 \pm 13$ & 10.87 & $4.0 \times 10^8$ & $2.4 \times 10^8$ & $6.8 \times 10^8$ & gas & \cite{gultekin09,vdmarel98} \\
N7457 & S0 & 14.0 & $67 \pm 3$ & 9.42 & $1.0 \times 10^7$ & $4.0 \times 10^6$ & $1.6 \times 10^7$ & stars & \cite{schulze11} \\
N7582 & SBab & 22.3 & $156 \pm 19$ & & $5.5 \times 10^7$ & $4.4 \times 10^7$ & $7.1 \times 10^7$ & gas & \cite{gultekin09,wold06} \\
U3789 $^{b,d}$ & SA(r)ab & 48.4 & $107^{+13}_{-12}$ &  & $1.08 \times 10^7$ & $1.03 \times 10^7$ & $1.14 \times 10^7$ & masers & \cite{kuo11} \\
\hline
\end{tabular}
\end{center}

\textbf{Notes:}  $D$ is the distance to the galaxy, assuming $H_0 = 70 \kms$ Mpc$^{-1}$.  $L_{V,\rm sph}$ is the $V$-band stellar luminosity of the galaxy's spheroidal component.  $\mbh ,_{\rm min}$ and $\mbh ,_{\rm max}$ are the upper and lower bounds of the 68\% confidence interval in $\mbh$.\\

$^a$  The literature contains a large number of estimates for the velocity dispersion of our Galaxy's bulge, using different kinematic tracers at different radii.  We use the radially averaged measurement of $\sigma = 103 \pm 20 \kms$ from Tremaine et al. (2002) \cite{tremaine02}. 

$^b$  G09 use the convention $H_0 = 70 \kms$ Mpc$^{-1}$ for all galaxies whose distances are derived from systemic velocity measurements.  To match this convention, we have adjusted the distance to several galaxies in our updated sample.  All distance adjustments yield corresponding adjustments to $\mbh$, $\mbh ,_{\rm min}$, and $\mbh ,_{\rm max}$, such that $\mbh \propto D$.  Our reported 68\% confidence intervals for $\mbh$ do not include uncertainties in the distance measurements.

$^c$  We derive an effective velocity dispersion of $315 \kms$ for IC 1459, using spatially resolved major-axis kinematics from Cappellari et al. (2002) \cite{cappellari02}, at radii of 0.8 - 25.1 arcseconds.  For $r < 0.8$ arcseconds, stellar motions are directly influenced by the central black hole.

$^d$  Maser-based black hole masses for several galaxies are presented in Greene et al. (2010) \cite{greene10} and Kuo et al. (2011) \cite{kuo11}.  We use the velocity dispersions presented in Greene et al. (2010).  For consistency with the rest of our sample, we use the black hole masses from Kuo et al. (2011), which agree with the values in Greene et al. (2010) but do not include distance uncertainties in the overall uncertainty for $\mbh$.  Braatz et al. (2010) \cite{braatz10} provide an updated distance and black hole mass for UGC 3789, which are consistent with the values we adopt from Kuo et al. (2011).

$^e$  Following G09, our sample includes two distinct measurements for NGC 1399 and also for NGC 5128.  We weight each of these measurements by 50\% when performing fits to $\mbh (\sigma)$ and $\mbh (L)$.  In Figure~\ref{fig:bhfits} we only plot the more recent measurement for each galaxy.

$^f$  We derive an effective velocity dispersion of $296 \kms$ for NGC 1399, using spatially resolved measurements from Graham et al. (1998) \cite{graham98} and Gebhardt et al. (2007) \cite{gebhardt07} at radii of 0.6 - 41 arcseconds.  For $r < 0.6$ arcseconds, stellar motions are directly influenced by the central black hole.

$^g$  NGC 3368 and NGC 3489 were not included in our sample until late in the editorial process, and so do not contribute toward the fits reported in the main letter or in Table 3.  Still, we view the reported black hole masses for NGC 3368 and NGC 3489 as valid measurements that should be included in future studies.  
For the velocity dispersions, we use the average values and outer limits of the various measurements reported by Nowak et al. (2010) \cite{nowak10}.
Including NGC 3368 and NGC 3489, the $\mbh - \sigma$ parameters ($\alpha$, $\beta$, $\epsilon_0$) are ($8.28 \pm 0.06$, $5.13 \pm 0.34$, 0.42) for 67 galaxies, ($8.38 \pm 0.06$, $4.57 \pm 0.36$, 0.36) for 46 early-type galaxies, and ($8.00 \pm 0.21$, $4.76 \pm 1.15$, 0.43) for 21 late-type galaxies.  These parameters are consistent with our reported values for the sample of 65 galaxies.

$^h$  The literature contains two inconsistent estimates of the stellar luminosity in NGC 3607: $M_V = -21.62$ in G09, and $M_V = -19.88$ in Lauer et al. (2007) \cite{lauer07}.  

$^i$ We derive an effective velocity dispersion of $270 \kms$ for NGC 3842, using measurements from GMOS and VIRUS-P at radii of 1.2 - 29.8 arcseconds.  For $r < 1.2$ arcseconds, stellar motions are directly influenced by the central black hole.

$^j$  For M87, we use the updated velocity dispersion of $324 \kms$ from Gebhardt et al. (2011) \cite{gebhardt11}, based on measurements at radii of 2.1 - 100 arcseconds.  For $r < 2.1$ arcseconds, stellar motions are directly influenced by the central black hole.

$^k$  The literature contains two inconsistent estimates of the bulge stellar luminosity in NGC 4564: $M_V = -19.60$ in G09, and $M_V = -20.26$ in Lauer et al. (2007) \cite{lauer07}.  

$^l$  We derive an effective velocity dispersion of $341 \kms$ for M60, using spatially resolved measurements from Pinkney et al. (2003) \cite{pinkney03} at radii of 2.2 - 44 arcseconds.  For $r < 2.2$ arcseconds, stellar motions are directly influenced by the central black hole.

$^m$  Our quoted value of $\mbh$ for NGC 4889 is the median of the 68\% confidence interval $0.7 - 3.4 \times 10^{10} \msun$.  We derive an effective velocity dispersion of $347 \kms$, using kinematics from Loubser et al. (2008) \cite{loubser08} at radii of 2.0 - 22.9 arcseconds.  The inner cutoff of 2.0 arcseconds excludes the asymmetric velocity dispersion peak and possible stellar torus.
\end{table*}


\end{document}